\documentclass{article}

\usepackage{arxiv}
\usepackage{float}
\usepackage[utf8]{inputenc} 
\usepackage[T1]{fontenc}    
\usepackage[hidelinks]{hyperref}
\usepackage{booktabs}       
\usepackage{amsfonts}       
\usepackage{nicefrac}       
\usepackage{microtype}      
\usepackage{graphicx}
\usepackage{doi}
\usepackage{amsmath}
\usepackage{listings}
\usepackage{xcolor}
\usepackage[numbers]{natbib}
\usepackage{enumitem}

\title{PRISM-UDE: Physics-Regularized Iterative Symbolic Modeling of 3nm FinFETs via Universal Differential Equation}

\author{
    \textbf{Pranavanath Balamurali} \\
    The University of Texas at Austin \\
    \texttt{pb27677@utexas.edu}
    \and
    \textbf{Prathamesh Dinesh Joshi} \\
    Vizuara AI Labs \\
    \texttt{prathamesh@vizuara.com}
    \and
    \textbf{Raj Abhijit Dandekar} \\
    Vizuara AI Labs \\
    \texttt{raj@vizuara.com}
    \and
    \\
    \textbf{Rajat Dandekar} \\
    Vizuara AI Labs \\
    \texttt{rajatdandekar@vizuara.com}
    \and
    \\
    \textbf{Sreedath Panat} \\
    Vizuara AI Labs \\
    \texttt{sreedath@vizuara.com}
}

\renewcommand{\undertitle}{Research Article}
\renewcommand{\shorttitle}{PRISM-UDE: Compact Model Synthesis for 3nm FinFETs}

\hypersetup{
pdftitle={PRISM-UDE: Automated Compact Model Synthesis for 3nm FinFETs via Physics-Regularized Symbolic Modeling},
pdfsubject={Electrical Engineering, Device Physics, Scientific Machine Learning},
pdfauthor={Pranavanath Balamurali},
pdfkeywords={3nm FinFET, PRISM-UDE, Universal Differential Equations, Symbolic Regression, Physics-Regularized ML, Log-MSE, SPICE Compact Modeling},
}

\begin{document}
\maketitle

\begin{abstract}
Compact transistor models are the mathematical backbone of circuit simulation. However, at advanced nodes such as 3\,nm, transport physics becomes too complex for traditional hand-derived equations to capture accurately. Purely data-driven neural surrogates, on the other hand, are numerically unstable inside circuit solvers and offer no physical insight into their own predictions. We introduce \textbf{PRISM-UDE} (Physics-Regularized Iterative Symbolic Modeling via Universal Differential Equations), a framework that embeds a small neural network inside a physics-based transistor model, using the network only to learn the transport behavior that the analytical baseline misses, rather than replacing the physics altogether. Once trained, this neural correction is distilled into a single, interpretable closed-form expression via symbolic regression, making the final model fully analytical and simulator-ready. Applied to a 3\,nm FinFET benchmark dataset, PRISM-UDE reduces prediction error more than sixfold (70.33\% to 11.01\%) relative to the standard physics-only baseline. The distilled expression preserves this accuracy almost exactly while eliminating the neural network entirely. We further validate the extracted expression directly inside a SPICE circuit simulator, confirming stable, physically consistent behavior under both static bias sweeps and dynamic switching conditions.
\end{abstract}


\section{Introduction}

Modern Electronic Design Automation (EDA) relies on compact models: highly optimized, physics-grounded mathematical surrogates that describe current-voltage (I-V) behavior of semiconductor transistors. These models serve as the fundamental bridge between physical manufacturing and circuit simulation. However, the gap between what compact models can describe and what the devices actually do is widening as transistors scale down to 3 nm and below \cite{moore2006cramming, lundstrom2006nanoscale, lundstrom2013near}. Gate lengths are now on the order of 10–20\,nm, and the effective channel is often sub-10\,nm \cite{lundstrom2013near}. In this regime, classical drift-diffusion equations become only a rough approximation: carrier mean free paths are comparable to the channel length, transport is partly ballistic, and strong vertical confinement changes subband structure and effective masses~\cite{lundstrom2006nanoscale, lundstrom2013near}.  Additionally, aggressive scaling of oxide thickness and supply voltage pushes the device close to quantum confinement limits, shifting threshold voltages and altering off-state behavior in ways that have no simple classical counterpart~\cite{ando1982electronic, duarte2015bsim, bae3nmfinfet}. At these nanoscale dimensions, short-channel effects become dominant. Drain-induced barrier lowering (DIBL) makes the threshold voltage strongly bias-dependent. Gate-induced drain leakage (GIDL) and band-to-band tunneling introduce non-negligible off-state currents~\cite{taur2021fundamentals}. The FinFET geometry itself,
while improving electrostatic control, adds modeling complexity through multi-fin variability, surface roughness on fin sidewalls, and fringing fields between adjacent structures~\cite{bae3nmfinfet, auth2012finfet}. These effects are strongly coupled and nonlinear, which makes it difficult to not only capture device behavior with a fixed set of closed-form expressions, but also interpret these equations.

The mathematical precision of these multi-decade current-voltage
($I$--$V$) trajectories represents a critical point of failure for modern
Electronic Design Automation (EDA) toolchains.  Circuit simulators rely on
iterative Newton-Raphson numerical solvers to evaluate designs that
contain billions of interconnected devices \cite{Nagel1975SPICE, kundert1995spice}.
Therefore, obtaining accurate predictions $I$--$V$ and absolute derivative
continuity is non-negotiable \cite{duarte2015bsim}.  Even
minor, localized anomalies or unphysical conductance discontinuities ($C^1$
or $C^2$ metrics) trigger gradient explosions that cause solver
divergence and crash the simulation pipeline.  Furthermore, because circuit
architects utilize these models to calculate nanoscale gate delays and power
dissipation matrices, any modeling discrepancy directly propagates into
fatal chip-level timing violations or severe power leakage \cite{chadha2009sta}.
At advanced process nodes, physical variation sources such as random dopant
fluctuations and line-edge roughness compound these challenges, making
statistical accuracy across corners equally important to worst-case
accuracy at nominal bias \cite{asenov1998variability, sano2002rdf}.
Ultimately, maximizing localized transport model precision minimizes structural variations during pre-fabrication analysis, serving as a critical step toward reliable early-stage device yield evaluation.
 
To address these modeling requirements, the current industry standard
relies on the Berkeley Short-channel IGFET Model for Common Multi-Gate
architectures (BSIM-CMG) \cite{dunga2008bsimcmg}.  This traditional
framework attempts to capture semiconductor physics through a vast,
hierarchically arranged network of empirical closed-form equations.
The BSIM-CMG model incorporates dedicated sub-modules for drift-diffusion
core transport, charge-based unified electrostatics, quantum mechanical
threshold shifts, gate-oxide tunneling, and layout-dependent geometric
effects \cite{dasgupta2020bsimcmg}.  However, maintaining
the fidelity of a BSIM model card at the 3\,nm node requires the
simultaneous calibration of a massive parameter space, frequently numbering hundreds of individual coefficients.
 
This parameter extraction workflow creates a serious engineering bottleneck. It demands multi-variable non-linear optimization across massive reference tables under varying bias, temperature, and geometric configurations: a process that is computationally expensive, highly technology-dependent, and heavily reliant on manual tuning by domain experts \cite{chen2025paramextract}. The scaling of the extraction problem at each new process node is non-trivial: the number of interdependent parameters grows, the non-convexity of the
optimization landscape deepens, and the tolerance budgets tighten as supply voltages drop \cite{leonhardt2015finfet}.  More fundamentally,
non-classical nanometer transport anomalies are often too complex to be
captured by rigid classical expressions. Forcing these empirical
formulations to fit multivariate data typically requires localized
mathematical smoothing approximations, which directly degrade numerical
stability, destroy physical interpretability, and compromise the solver
convergence vital to downstream circuit design \cite{mcandrew2015bestpractices}.

To automate these labor-intensive engineering pipelines, recent research in Scientific Machine Learning (SciML) has focused on leveraging deep neural networks to extract non-linear transistor behavior directly from high-fidelity reference simulation data~\cite{woo2022mlcompact, novkin2026kan, mamun2025review}. While unconstrained, purely data-driven surrogate configurations achieve exceptional accuracy within their interpolation domains \cite{zhang2017ann}, they suffer from three limitations that inhibit direct integration into standard EDA environments:

\begin{enumerate}
\item \textbf{Loss of Derivative Continuity:} Purely data-driven neural fields lack structural physical regularizers, frequently generating hidden slope non-monotonicities or non-physical negative differential resistances ($\partial I_{ds}/\partial V_{ds} < 0$) when evaluated outside narrow training boundaries \cite{mcandrew2015bestpractices}.
\item \textbf{Fatal SPICE Non-Convergence:} When integrated into a numerical circuit solver, localized derivative anomalies disrupt the iterative Newton-Raphson matrix loop, causing solver stagnation or matrix underflow \cite{kundert1995spice}.
\item \textbf{Computational Inefficiency:} Deploying neural surrogates in iterative solvers requires evaluating multi-layer matrix operations and non-linear activation functions at every solver step \cite{hutchins2022generalized}. Compared to compiled, closed-form algebraic expressions, this operational burden creates runtime evaluation latency that hinders scaled adoption in commercial SPICE simulators.
\end{enumerate}

To circumvent these limitations, Physics-Informed Neural Networks (PINNs)~\cite{raissi2019physics} augment training procedures by penalizing partial differential equation residuals within the loss objective. However, for compact modeling applications, the objective is not to solve a global continuum field, but rather to construct a computationally efficient, stable analytical formulation that captures high-dimensional residuals relative to an industry-standard benchmark. This motivates the deployment of the Universal Differential Equation (UDE) paradigm~\cite{rackauckas2021ude, rackauckas2017sciml}. In a UDE framework, a deep neural network is embedded directly as a sub-term within an established macroscopic differential system rather than replacing it, restricting the learned degrees of freedom to the physically uncharacterized transport residual.
We term this framework \textbf{PRISM-UDE} (Physics-Regularized Iterative Symbolic Modeling via Universal Differential Equations). Its central design philosophy inverts the traditional empirical model extraction approach. Instead of enumerating each micro-scale transport mechanism and introducing an isolated mathematical parameter for each, PRISM-UDE starts from a structurally sound macroscopic drift-diffusion scaffold that is continuous at the boundary level. It then embeds a neural network directly inside that scaffold as a residual correction term ($\Delta\mu_{\text{NN}}$), which absorbs only the transport physics the scaffold cannot express. Finally, symbolic regression is applied to distill the trained network back into a single closed-form algebraic expression. As a result, the final deployed model contains no neural network at all.
Since the physical baseline already handles fundamental conservation laws and strict boundary constraints, the neural network correction term ($\Delta\mu_{\text{NN}}$) operates within a low-dimensional correction space, minimizing overparameterization, regularizing gradients, and making the neural field highly amenable to evolutionary symbolic regression. All training, optimization, and validation configurations are executed using the standardized MOSFET Electrical Simulation Dataset (MESD)~\cite{zhang2024mesd}, a high-fidelity benchmark repository of standardized multi-decade electrical simulation profiles generated via foundry-calibrated BSIM engines across multi-temperature and physical scaling boundaries. We can explicitly focus PRISM-UDE as an MESD-to-expression synthesis pipeline targeting the 3nm FinFET node. This establishes a controlled, highly reproducible environment. From there, we can evaluate whether automated model synthesis is truly feasible against an established numerical ground truth.

This work focuses strictly on the synthesis, mathematical regularization, and dynamic validation of continuous, non-linear transport equations ($I$--$V$ characterization sweeps). While evaluating terminal charge conservation properties ($C$--$V$ transcapacitances) is vital for dynamic delay profiling in multi-stage logic architectures, the automated mapping of displacement charges via charge-conserving UDE formulations is designated as a separate track for future work \cite{ward1978charge}. Consequently, the resulting symbolic expressions demonstrate the feasibility of physics-augmented machine learning for SPICE-compatible transport-model prototyping rather than representing a fully realized, industrial-grade commercial PDK surrogate.

Summary of our contributions:
\begin{itemize}
\item \textbf{Unified Multi-Decade Charge-Sheet Baseline:} A hybrid compact-modeling framework combining a unified physical inversion charge ($Q_{inv}$) formulation and a Caughey--Thomas mobility baseline to serve as a strict numerical regularizer, enforcing definitive boundary constraints ($\lim_{V_{ds}\to0} I_{ds} = 0$) and continuous derivative trajectories inside SPICE.

\item \textbf{Decoupled Two-Stage Optimization Loop:} An optimization pipeline that uses Particle Swarm Optimization (PSO) to anchor macroscopic baseline parameters before unlocking the neural network, preventing neural capacity from absorbing known physical bounds.

\item \textbf{Multi-Decade Logarithmic Loss Formulation:} A continuous $\log_{10}$-space
mean-squared-error objective that equalizes gradient weight across the seven-decade
dynamic range of the transistor current ($10^{-10}$~A to $10^{-3}$~A), mitigating
on-state optimization bias. Regularization is structural rather than penalty-based: the
physical scaffold, not an explicit weight-norm term, constrains the neural residual.

\item \textbf{Evolutionary Symbolic regression:} Application of genetic programming~\cite{koza1994gp, cranmer2023pysr} to compress the $\tanh$-based neural field into a compact closed-form algebraic expression ($\Delta\mu_{\text{extracted}}$), showing less than 1.4 percentage points of device-level MRE degradation relative to the neural model.

\item \textbf{End-to-End Dynamic SPICE Verification:} Implementation of the extracted analytical model in LTspice using a behavioral source paired with a $\mathcal{C}^\infty$-continuous smooth-absolute approximation to eliminate non-smooth derivative discontinuities \cite{nocedal2006numerical}, demonstrating numerical convergence across DC sweeps and high-frequency transient square-wave switching testbenches.
\end{itemize}

\section{Related Work}

\subsection{Compact Modeling Standards and Transistor Datasets}

The Berkeley Short-channel IGFET Model for Common Multi-Gate architectures (BSIM-CMG) represents the global industry standard for compact modeling of advanced multi-gate technologies and serves as the primary physical benchmark against which modern device research is evaluated. Developed by the BSIM group at UC Berkeley, it incorporates drift-diffusion core transport, charge-based electrostatics, quantum mechanical threshold shifts, gate-oxide tunneling, and an extensive matrix of layout-dependent effect (LDE) equations \cite{dunga2008bsimcmg, dasgupta2020bsimcmg}.

Alternatively, the MIT Virtual Source Model (MVS) provides an elegant semi-analytical formulation grounded in quasi-ballistic source-injection physics \cite{khakifirooz2009mvs, rakheja2015mvs}. While MVS excels at minimizing parameter counts in ultra-scaled channels where ballistic fractions approach unity, it sacrifices structural generality across multi-temperature scaling nodes.

To bridge the gap between complex analytical standards and data-driven architectures, modern device characterization increasingly relies on standardized benchmark repositories. The MOSFET Electrical Simulation Dataset (MESD) developed by Zhang et al.~\cite{zhang2024mesd} established a standardized, comprehensive multi-modal infrastructure for compact model evaluation. The dataset provides vast collections of high-fidelity $I$--$V$ and $C$--$V$ curves spanning multiple advanced process nodes down to 3nm dimensions under strict structural and temperature variations, serving as the foundational ground truth for the framework developed in this manuscript.

\subsection{Machine Learning Surrogates for Semiconductor Devices}

The optimization of data-driven surrogates to map multi-decade transistor characteristics has been an active area of research. Previous literature has demonstrated the training of deep multi-layer perceptrons on reference transistor datasets, achieving sub-1\% interpolation bounds but noting fatal errors when evaluated outside training coordinates \cite{novkin2026kan, zhang2017ann}. Kim et al.~\cite{kim2022simulator} explored gradient-boosted trees to enhance computational extraction efficiency, while Guven et al.~\cite{guven2025surrogate} discuss regression models for rapid design-space exploration.

To enforce physical boundaries, Physics-Informed Neural Networks (PINNs)~\cite{raissi2019physics} insert numerical derivative penalties into the loss formulation. However, these architectures output active neural network weight matrices that require unoptimized computational runtime wrappers. Consequently, none of these existing methodologies generates a closed-form, purely analytical mathematical expression that can be compiled directly into a compact model netlist card without structural simulator modification \cite{mcandrew2015bestpractices}.

\subsection{Prior Applications of Hybrid Physical–Neural Modeling}

The Universal Differential Equation (UDE) framework was formalized by Rackauckas et al.~\cite{rackauckas2021ude} as a unified paradigm for augmenting incomplete mathematical models with data-driven neural operators. UDE frameworks have successfully mapped complex physical behaviors in atmospheric chemistry closure models, stiff biochemical kinetic pipelines, and turbulent fluid mechanics systems. By constraining the deep neural layer to a localized, low-dimensional functional correction zone defined by surrounding differential loops, UDEs exhibit superior sample efficiency and extrapolation compared to unconstrained surrogates.

To restore interpretability to these hybrid structures, Cranmer et al.~\cite{cranmer2023pysr} and Udrescu et al. \cite{udrescu2020aifeynman} introduce a multi-stage approach where symbolic regression is applied to a pre-trained neural network to extract clean analytical equations. PySR provides a high-performance, GPU-accelerated evolutionary genetic algorithm optimized for searching wide Pareto-front algebraic spaces.

The integration of a log-space Mean Squared Error objective to balance multi-decade optimization surfaces is well-established within the literature~\cite{zhang2017ann}. However, the application of this combined pipeline, using a UDE to regularize 3nm transport boundaries and distilling the learned neural field into a $\mathcal{C}^\infty$ SPICE-compatible algebraic equation, represents a novel contribution to the field of automated compact model synthesis.

\section{Methodology}

\subsection{Dataset Curation and Benchmark Configuration}

To train, optimize, and evaluate the hybrid UDE framework under standardized and highly reproducible conditions, we leverage the Metal-Oxide-Semiconductor Field-Effect Transistor (MOSFET) Electrical Simulation Dataset (MESD)~\cite{zhang2024mesd}. MESD serves as an extensive open-source benchmark featuring high-fidelity steady-state current--voltage ($I\text{--}V$) and dynamic capacitance--voltage ($C\text{--}V$) characteristics simulated across multiple foundry Process Design Kits (PDKs) spanning technology nodes from 3~nm to 350~nm. The dataset captures multi-dimensional sweep trajectories—including varying bias voltages, operating temperatures, and device geometries—generated via BSIMs in industry-standard EDA engines (e.g., Cadence Spectre and Synopsys HSPICE).

For the 3nm multi-gate transport synthesis executed in this work, we isolate the physics-calibrated \texttt{N3A} 3nm FinFET process partition from the MESD. To bypass foundry non-disclosure restrictions and prevent process property leakage, all foundry-specific nomenclature is anonymized within the dataset structural layers while preserving the precise physical couplings and short-channel anomalies characteristic of the 3nm node. The selected \texttt{N3A} structural matrix provides a comprehensive multi-dimensional sweeping space across varying terminal biases ($V_{gs}$, $V_{ds}$), environmental boundaries (Temperature $T$), and device geometries defined by multi-fin arrays ($N_{fin}$) and sub-micron gate lengths ($L$), as explicitly detailed in Table~\ref{tab:mesd_structure}.

\begin{table}[H]
\centering
\caption{Structural Fields and Feature Dimensions of the Isolated MESD \texttt{N3A} FinFET Partition}
\label{tab:mesd_structure}
\small
\begin{tabular}{lll}
\toprule
\textbf{Field Entity} & \textbf{Physical / Modeling Description} & \textbf{Data Representation \& Units} \\
\midrule
\texttt{PDK}       & Anonymized Process Design Kit Identification Tag  & String (\texttt{"N3A"}) \\
\texttt{Node}      & Target lithography process minimum feature boundary & Integer ($3$~nm) \\
\texttt{Device}    & Multi-gate device structural designation            & String (\texttt{"NMOS"}) \\
\texttt{Corner}    & Process variation boundaries                        & String (\texttt{"TT"}) \\
\texttt{Temp}      & Environmental validation temperature matrix         & Integer ($-40^\circ\text{C}$ to $125^\circ\text{C}$) \\
\texttt{L}         & Effective drawn physical gate length                & Integer ($15$~nm) \\
\texttt{Nfin}      & Structural parallel multi-fin configuration count   & Discrete Integers ($1, 2, 4, 8, 16$) \\
\texttt{Vds}       & Controlled lateral drain-to-source voltage bias     & Discrete Floats ($0.0$~V to $0.7$~V, step $0.1$~V) \\
\texttt{Vgs}       & Swept vertical gate-to-source voltage bias grid     & Discrete Floats ($0.0$~V to $0.7$~V, step $0.1$~V) \\
\texttt{Ids}       & Ground-truth reference channel drain current        & List of Floats (Amperes) \\
\bottomrule
\end{tabular}
\end{table}

The dataset partition is compiled into structured JSON configurations containing $175$ independent device multi-bias trajectories. The dataset was split using a stratified randomized partitioning scheme: $140$ trajectories ($80\%$) were allocated to the Stage 1 and Stage 2 optimization loops, while the remaining $35$ trajectories ($20\%$) were strictly withheld to serve as an independent evaluation partition for testing the generalization limits of the distilled symbolic compact equations.

\begin{figure}[H]
\centering
\includegraphics[width=\textwidth]{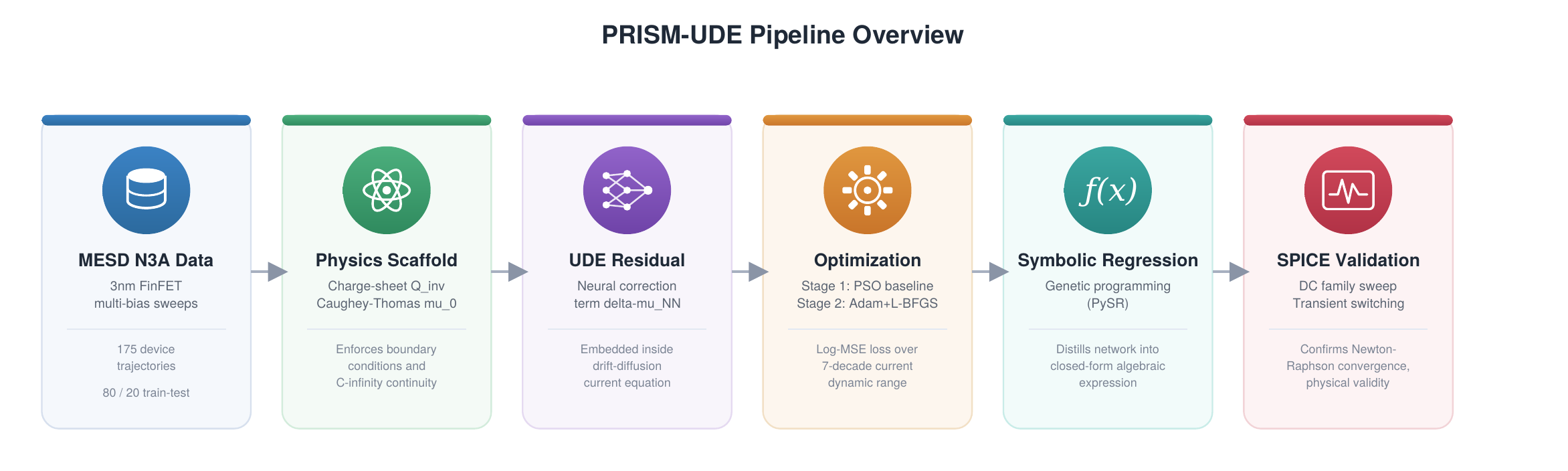}
\caption{End-to-end PRISM-UDE pipeline: a physics-grounded transport scaffold constrains a neural residual correction, which is jointly optimized in two stages, distilled into a closed-form expression via symbolic regression, and validated directly inside a SPICE circuit simulator.}
\label{fig:methodology_overview}
\end{figure}

\subsection{Theoretical Framework of Universal Differential Equations}
\label{sec:ude_theory}
 
A Universal Differential Equation (UDE) is a dynamical system in which one
or more unknown or difficult-to-specify terms are replaced by a universal
function approximator—typically a neural network—while the rest of the
system retains its known analytical structure \cite{rackauckas2021ude}.
The general form for a system of differential equations is:
 
\begin{equation}
\frac{d\mathbf{u}}{dt} = f\!\left(\mathbf{u},\, t,\, U_\theta(\mathbf{u}, t)\right),
\quad \mathbf{u}(t_0) = \mathbf{u}_0
\label{eq:ude_general}
\end{equation}
 
where $\mathbf{u}$ is the state vector, $f$ encodes the known physics of the
system, and $U_\theta(\mathbf{u}, t)$ is a neural network with trainable
parameters $\theta$ that approximates the unknown or missing physical term.
The key distinction from a pure neural ODE \cite{chen2018neuralode} is that
$f$ is \emph{not} entirely replaced by the network; instead, the network
occupies only the gap between the known model and the true system behavior.
This structural constraint has two important consequences: (i) the network
only needs to learn a low-dimensional residual, dramatically reducing the
effective parameter count and improving sample efficiency; and (ii) the
known structure of $f$ enforces physical boundary conditions and conservation
laws automatically, regardless of what $\theta$ converges to.
 
For compact device modeling, the ``differential system'' is the set of
transport equations governing the channel current $I_{ds}$.  The known
physics—drift-diffusion transport, unified charge-sheet electrostatics,
and Caughey--Thomas velocity saturation—forms $f$.  The unknown term is the
high-dimensional residual mobility correction that accounts for
quantum confinement, ballistic velocity overshoot, multi-gate fringing
fields, and phonon-scattering anomalies that the classical baseline cannot
capture.  We write the effective carrier mobility as an additive UDE:
 
\begin{equation}
\mu_{\text{eff}}(\mathbf{x}) = \underbrace{\mu_0(\mathbf{x})}_{\text{known physics}} +
\underbrace{U_\theta(\mathbf{x})}_{\text{neural residual } \Delta\mu_{\text{NN}}}
\label{eq:ude_mobility}
\end{equation}
 
where $\mathbf{x} = [V_{gs}, V_{ds}, \text{TRatio}, N_{fin}, L]$ is the
input feature vector (defined explicitly in Section~\ref{sec:scaffolding}). To ensure $\mu_{\text{eff}} > 0$ across all training coordinates and prevent solver instability during optimization, the additive sum $\mu_0 + U_\theta$ is passed through a smooth $\mathcal{C}^\infty$ softplus positivity transformation during evaluation (Section~\ref{sec:positivity}, Eq.~\ref{eq:mu_positive}).

The total drain current then inherits the structure of the physical baseline
while the network correction $\Delta\mu_{\text{NN}}$ absorbs only what the
analytical model misses:
 
\begin{equation}
I_{ds}(\mathbf{x}) = W \cdot Q_{inv}(\mathbf{x}) \cdot
\bigl[\mu_0(\mathbf{x}) + U_\theta(\mathbf{x})\bigr] \cdot E_{\text{eff}}(V_{ds})
\label{eq:ude_ids}
\end{equation}
 
Because $\mu_0$ already enforces $I_{ds} \to 0$ as $V_{ds} \to 0$ and
monotone subthreshold behavior, $U_\theta$ is constrained to a
physically plausible correction space even before training begins.
This prevents the network from absorbing macroscopic physical trends and
limits it to isolating only the high-dimensional transport anomalies
present in the 3\,nm reference data.
 
The core engineering philosophy of this work addresses the key limitations of contemporary compact device modeling: the intensive, multi-variable manual parameter optimization required by traditional empirical formulations, and the physically unconstrained, numerically volatile behavior inherent to pure black-box machine learning approaches.  Pure machine learning architectures, such as
unconstrained PINNs or standard Multi-Layer Perceptrons (MLPs), treat
terminal electronic characteristics as generic, high-dimensional
surface-fitting landscapes.  When inserted into the mathematical core of
an EDA matrix solver, they routinely manifest severe non-physical
artifacts, including localized negative differential resistances
($\partial I_{ds}/\partial V_{ds} < 0$), spurious non-monotonic
transconductance profiles, or hidden slope discontinuities.  When
processed by the simulator's iterative Newton-Raphson engine, these
anomalies cause immediate local derivative instability, leading to
matrix underflow or non-convergence errors.
 
UDE frameworks have successfully resolved analogous residual-learning
problems in other physical domains, including systems biology metabolic
networks \cite{ude_sysbio2025} and glacier ice-flow dynamics
\cite{bolibar2023ude}, demonstrating superior sample efficiency and
extrapolation relative to unconstrained surrogates in each case.  Their
application to semiconductor compact modeling, where a precise analytical
structure already exists and the goal is to learn the gap between that
structure and reality, is a natural and well-motivated extension \cite{mdnn2024tcad}.

\subsection{Macroscopic Transport Framework and Unified Electrostatics}
\label{sec:scaffolding}
 
The foundation of the structural baseline begins with a macroscopic
drift-diffusion transport equation defining the steady-state
drain-to-source current flow \cite{tsividis1999mosfet, arora1993mosfet}:
\begin{equation}
I_{ds}(V_{gs}, V_{ds}, \text{TRatio}) =
  W \cdot Q_{inv}(V_{gs}, V_{ds}, \text{TRatio})
           \cdot \mu_{\text{eff}}(V_{gs}, V_{ds}, \text{TRatio})
           \cdot E_{\text{eff}}(V_{ds})
\label{eq:transport_core}
\end{equation}
where:
\begin{center}
\small
\renewcommand{\arraystretch}{1.15}
\begin{tabular}{lll}
\toprule
\textbf{Symbol} & \textbf{Description} & \textbf{Units} \\
\midrule
$V_{gs}$ & Gate-to-source voltage & V \\
$V_{ds}$ & Drain-to-source voltage & V \\
$\text{TRatio} = T/T_{\text{ref}}$ & Normalized temperature ratio ($T_{\text{ref}}=300.15$\,K) & -- \\
$W = N_{fin}\cdot W_{fin}$ & Total gate width across all fins & m \\
$L$ & Effective channel length ($15$\,nm, \texttt{N3A}) & m \\
$Q_{inv}$ & Inversion layer charge density & C/m$^2$ \\
$\mu_{\text{eff}}$ & Effective carrier mobility & m$^2$/V$\cdot$s \\
$E_{\text{eff}}$ & Effective lateral electric field & V/m \\
\bottomrule
\end{tabular}
\end{center}
 
To guarantee accurate physical matching across both the exponential
subthreshold regime and the strong-inversion linear/saturation regions,
the inversion charge density $Q_{inv}$ is modeled via a mathematically
unified charge-sheet formulation:
\cite{brews1978chargesheet}:
\begin{equation}
Q_{inv}(V_{gs}, V_{ds}, \text{TRatio}) =
  C_{ox} \cdot V_{gst}
  \cdot \left[1 - \frac{V_{ds}}{2\,(V_{gst} + 2\,v_t)}\right]
\label{eq:qinv}
\end{equation}
where $C_{ox}$ (F/m$^2$) is the gate dielectric capacitance per unit
area: the ratio of the gate dielectric permittivity to its physical
thickness, relating gate voltage to induced charge.  The bracketed
factor captures the linear reduction in inversion charge from source to
drain due to the applied $V_{ds}$.  This unified charge profile ensures
that the baseline current model remains continuous and infinitely
differentiable through the entire subthreshold-to-strong-inversion
transition zone.

The effective gate drive voltage
$V_{gst}$ (V), the voltage above threshold that drives strong-inversion
charge, is regularized across the threshold voltage $V_{th}$ (V) using a
smooth unified electrostatic function \cite{brews1978chargesheet}:
\begin{equation}
V_{gst} = 2 m v_t \ln\!\left(1 + \exp\!\left(\frac{V_{gs} - V_{th}}{2 m v_t}\right)\right)
\end{equation}
where:
\begin{center}
\small
\renewcommand{\arraystretch}{1.15}
\begin{tabular}{lll}
\toprule
\textbf{Symbol} & \textbf{Description} & \textbf{Units} \\
\midrule
$v_t = k_BT/q$ & Thermal voltage ($\approx 25.85$\,mV at $T_{\text{ref}}$) & V \\
$m$ & Subthreshold ideality (body-effect) factor & -- \\
\bottomrule
\end{tabular}
\end{center}
 
At 3\,nm technology nodes, the threshold voltage $V_{th}$ cannot be treated
as a static scalar constant.  We explicitly formulate a non-linear threshold
voltage expression incorporating drain-induced barrier lowering (DIBL) and
structural variations across process boundaries \cite{troutman1979dibl}:
\begin{equation}
V_{th}(V_{ds}, N_{fin}) = V_{th0} - \text{DIBL} \cdot V_{ds}
                           + \Delta V_{th,fin} \cdot (N_{fin} - 1)
\end{equation}
where:
\begin{center}
\small
\renewcommand{\arraystretch}{1.15}
\begin{tabular}{lll}
\toprule
\textbf{Symbol} & \textbf{Description} & \textbf{Units} \\
\midrule
$V_{th0}$ & Zero-bias threshold, single fin & V \\
$\text{DIBL}$ & Barrier-lowering coefficient ($\approx 0.1$\,V/V typical) & V/V \\
$N_{fin}$ & Number of parallel fins & -- \\
$\Delta V_{th,fin}$ & Per-fin threshold shift coefficient & V \\
\bottomrule
\end{tabular}
\end{center}

The effective lateral electric field $E_{\text{eff}}$ accounts for
short-channel field non-uniformity near the drain terminal via a
characteristic velocity saturation length parameter $\ell$ (m):
\begin{equation}
    E_{\text{eff}}(V_{\text{ds}}) = \frac{V_{\text{ds}}}{\sqrt{L^2 + \left(\frac{V_{\text{ds}}}{E_{\text{sat}}}\right)^2}}
\end{equation}
where $E_{sat} = 2\,v_{sat}/\mu_{00}$ (V/m) is the critical field at which carrier velocity saturates, $v_{sat}$ (m/s) is the carrier saturation velocity, and $\mu_{00}$ (m$^2$/V$\cdot$s) is the baseline low-field mobility at $T_{\text{ref}}$.  Note that $E_{\text{eff}}$ is a field rather than a voltage: the channel-length
normalization enters through the $L^{2}$ term in its denominator, so the transport
equation is written with a $W$ prefactor and not $W/L$. Equations~\eqref{eq:ude_ids}
and~\eqref{eq:transport_core} therefore describe the same current, and this is the form
evaluated in both the training code and the SPICE behavioral source.
 
The underlying numerical regularizer for the effective mobility is a
modified Caughey--Thomas velocity saturation model \cite{caughey1967mobility}:
\begin{equation}
\mu_{0}(V_{ds}, \text{TRatio}) =
  \frac{\mu_{00}\,\text{TRatio}^{\alpha_\theta}}
       {\left[\,1 + \left(\dfrac{\mu_{00}\,\text{TRatio}^{\alpha_\theta}\,|E_{\text{eff}}(V_{ds})|}{v_{sat}}\right)^{\!\beta}\,\right]^{1/\beta}}
\label{eq:ct}
\end{equation}
where:
\begin{center}
\small
\renewcommand{\arraystretch}{1.15}
\begin{tabular}{lll}
\toprule
\textbf{Symbol} & \textbf{Description} & \textbf{Units} \\
\midrule
$\mu_{00}$ & Low-field mobility at $T_{\text{ref}}$ & m$^2$/V$\cdot$s \\
$\alpha_\theta$ & Temperature exponent ($\approx -1.5$ to $-2.2$ in Si) & -- \\
$v_{sat}$ & Carrier saturation velocity ($\approx 10^5$\,m/s in Si) & m/s \\
$\beta$ & Saturation-abruptness exponent & -- \\
\bottomrule
\end{tabular}
\end{center}
 
This macroscopic baseline is not expected to independently capture
the advanced transport physics inherent to 3\,nm technology nodes, such
as subband quantum mechanical confinement, severe DIBL, gate fringing
field variations, or ballistic velocity overshoot.  Instead, it operates
as a rigid \textit{mathematical anchor} that enforces definitive physical
boundary conditions directly into the solver's matrix.  Most critically,
it guarantees:
\begin{equation}
\lim_{V_{ds} \to 0} I_{ds}(V_{gs}, V_{ds}, \text{TRatio}) = 0
\label{eq:zero_bias}
\end{equation}
and mathematically forces continuous, strictly monotonic, well-behaved
derivative trajectories at zero-bias conditions, preventing floating nodes
or unconstrained optimization states during transient switching events.

\subsection{Neural Residual Embedding and UDE Formulation}

With the boundary regularizer rigidly established, the complete true effective carrier mobility $\mu_{\text{eff}}$ is formulated as an additive composition of the physical Caughey--Thomas baseline and a deep neural network residual correction field ($\Delta\mu_{\text{NN}}$):
\begin{equation}
\mu_{\text{eff}}(V_{gs}, V_{ds}, \text{TRatio}) = \mu_{0}(V_{ds}, \text{TRatio}) + \text{DNN}(V_{gs}, V_{ds}, \text{TRatio}, N_{fin}, L; \theta)
\end{equation}
Note that while the total effective mobility $\mu_{\text{eff}}$ is also a function of vertical gate bias $V_{gs}$, the baseline physical mobility $\mu_0(V_{ds}, \text{TRatio})$ depends strictly on lateral drain field $E_{\text{eff}}(V_{ds})$ and temperature ratio $\text{TRatio}$. All vertical gate-field degradation and screening effects are absorbed by the neural residual field $\text{DNN}(\mathbf{x}; \theta)$. Here, $\theta$ denotes the complete vector of optimized internal weights and biases. The neural network accepts a five-dimensional normalized input vector:
\begin{equation}
\mathbf{x} = \begin{bmatrix} V_{gs}/V_{dd} \\ V_{ds}/V_{dd} \\ \text{TRatio} \\ N_{fin}/N_{fin,max} \\ L/L_{norm} \end{bmatrix}
\end{equation}
where $V_{dd} = 1.2$~V is the fixed voltage normalization constant applied to the bias
inputs. It is a scaling convention inherited from the feature-normalization stage, not
an operating supply rail: the MESD \texttt{N3A} bias sweep spans $0$--$0.7$~V, so the
normalized inputs occupy $[0, 0.583]$. $N_{fin,max}$ and $L_{norm}$ normalize structural
geometry across the training domain.

The feed-forward propagation through the network is governed by sequential tensor operations:
\begin{align}
\mathbf{h}^{(1)} &= \sigma\left(\mathbf{W}^{(1)}\mathbf{x} + \mathbf{b}^{(1)}\right) \\
\mathbf{h}^{(l)} &= \sigma\left(\mathbf{W}^{(l)}\mathbf{h}^{(l-1)} + \mathbf{b}^{(l)}\right), \quad \text{for } l = 2, 3, \dots, L-1 \\
\text{DNN}(\mathbf{x}; \theta) &= \mathbf{W}^{(L)}\mathbf{h}^{(L-1)} + \mathbf{b}^{(L)}
\end{align}
The network uses an input layer ($D_{in}=5$), two hidden layers with $H_1 = H_2 = 64$ nodes, and a single linear output neuron ($D_{out}=1$) generating the $\Delta\mu_{\text{NN}}$ correction value.

To guarantee that the structural modifications do not introduce sharp slope discontinuities or non-differentiable points into the solver's Jacobian matrix, the activation function $\sigma$ must possess a minimum continuity class of $\mathcal{C}^2$. Standard Rectified Linear Unit (ReLU) functions are strictly barred due to their non-differentiable corner anomalies at zero, which cause numerical chatter and convergence failures in SPICE. Instead, a smooth hyperbolic tangent ($\tanh$) operator is utilized uniformly across all hidden layers:
\begin{equation}
\sigma(z) = \tanh(z) = \frac{e^z - e^{-z}}{e^z + e^{-z}}
\end{equation}

\subsection{Positivity Guards and Domain Constraints}
\label{sec:positivity}

The additive UDE of Eq.~\eqref{eq:ude_mobility} places no sign constraint on
$U_\theta$, so an untreated network output can in principle drive
$\mu_{\text{eff}}$, and hence $I_{ds}$, negative. Because the training objective in
Eq.~\eqref{eq:log_loss} evaluates $\log_{10}(I_{\text{sim}}+\epsilon)$, a negative
argument would be undefined and the gradient would be lost. Three
$\mathcal{C}^{\infty}$ guards are therefore applied inside the forward model, before
the loss is evaluated and before the expression is emitted to SPICE. All three are
smooth by construction, so none of them introduces a discontinuity into the solver
Jacobian.

\paragraph{(i) Effective drain-bias limiter.}
The drain bias entering the charge and field terms is limited to the saturation bias
through a smooth minimum rather than a hard \texttt{min}$(\cdot)$ conditional:
\begin{equation}
V_{ds,\text{eff}} = \frac{V_{ds}\,V_{dsat}}{\left(V_{ds}^{2}+V_{dsat}^{2}\right)^{1/2}},
\qquad
V_{dsat} = \frac{V_{gst}\,E_{\text{sat}}L}{V_{gst}+E_{\text{sat}}L}
\label{eq:vdseff}
\end{equation}
This reproduces $V_{ds,\text{eff}}\!\to\!V_{ds}$ in the linear region and
$V_{ds,\text{eff}}\!\to\!V_{dsat}$ deep in saturation, while remaining differentiable
to all orders at the knee. It also keeps the bracketed charge-reduction factor of
Eq.~\eqref{eq:qinv} strictly positive, preventing the unphysical charge sign reversal
that the unlimited $V_{ds}$ form admits beyond pinch-off.

\paragraph{(ii) Mobility positivity transformation.}
The sum of the physical baseline and the neural residual is mapped onto the positive
real line by a softplus transformation with a fixed scale $s$:
\begin{equation}
\mu_{\text{eff}} = s\,\ln\!\left[1+\exp\!\left(\frac{\mu_{0}+\Delta\mu_{\text{NN}}}{s}\right)\right],
\qquad s = 10^{-4}~\text{m}^{2}\text{V}^{-1}\text{s}^{-1}
\label{eq:mu_positive}
\end{equation}
For $\mu_{0}+\Delta\mu_{\text{NN}} \gg s$ the transformation is the identity to within
floating-point precision, so it does not distort the fitted mobility anywhere in the
strong-inversion regime; it acts only where the residual would otherwise overwhelm the
baseline and force a sign change. Unlike a clipped
$\max(\mu_{0}+\Delta\mu_{\text{NN}},\,0)$, it has no derivative discontinuity and
therefore does not stall the L-BFGS phase.

\paragraph{(iii) Current floor at the loss boundary.}
Finally, the modelled current presented to the objective is floored at the same
$\epsilon$ that appears inside the logarithm:
\begin{equation}
I_{\text{sim}} = \tfrac{1}{2}\left[I_{ds} + \sqrt{I_{ds}^{2} + \epsilon^{2}}\,\right],
\qquad \epsilon = 10^{-15}~\text{A}
\label{eq:i_floor}
\end{equation}
which is a smooth rectifier: it passes $I_{ds}$ unchanged whenever
$I_{ds}\gg\epsilon$ and saturates to $\mathcal{O}(\epsilon)$ otherwise.

Guards (i)--(iii) are the reason the zero-bias boundary condition of
Eq.~\eqref{eq:zero_bias} survives training. They are not post-hoc corrections applied
to the plotted curves: they are part of the forward model that the optimizer sees, and
the same three operations are reproduced verbatim in the SPICE behavioral source of
Section~\ref{sec:spice_impl}, so the deployed expression and the trained model are
numerically identical.

\subsection{Two-Stage Decoupled Multi-Decade Logarithmic Optimization}

Advanced 3nm FinFET devices exhibit an aggressive multi-decade operational continuum. The drain current $I_{ds}$ spans over seven orders of magnitude between the deep subthreshold off-state leakage regime ($10^{-10}$~A) and the strong inversion on-state saturation drive regime ($10^{-3}$~A). If a standard Mean Squared Error (MSE) loss function is deployed, optimization gradients will be completely dominated by the high-magnitude on-state saturation region, while massive multi-decade discrepancies in the subthreshold leakage domain are treated as negligible noise.

To enforce uniform numerical precision across both operational extremes, we use a custom multi-decade $\log_{10}$-space objective:
\begin{equation}
\mathcal{L}_{\text{log}}(\theta, \mathbf{\Phi}) = \frac{1}{N} \sum_{i=1}^{N} \left| \log_{10}\left(I_{\text{sim}, i} + \epsilon\right) - \log_{10}\left(|I_{\text{MESD}, i}| + \epsilon\right) \right|^2
\label{eq:log_loss}
\end{equation}
where $\epsilon = 10^{-15}$~A provides numerical regularization near zero-bias to prevent logarithmic singularities, $\mathbf{\Phi}$ represents the macroscale baseline physical parameter set, and $\theta$ encapsulates the neural network weights.No explicit $L_2$ weight-decay term is added to Eq.~\eqref{eq:log_loss}. The residual
is regularized structurally: the physical scaffold fixes the macroscopic trends during
Stage 1, so the network is confined to a low-dimensional correction space and the
capacity that a weight penalty would otherwise have to suppress is never released.

The resulting loss surface is highly non-convex, so a decoupled two-stage optimization routine is executed:

\subsubsection{Stage 1: Macroscopic Baseline Initialization via PSO}

Before unlocking the neural network, the framework locks $\theta = \mathbf{0}$ and executes a multi-variable extraction targeting the baseline electrostatic parameters $\mathbf{\Phi} = \{V_{th0}, m, \text{DIBL}, \Delta V_{th,fin}\}$ using Particle Swarm Optimization (PSO). This gradient-free global heuristic approach \cite{kennedy1995pso} anchors the subthreshold swing, threshold boundaries, and linear DIBL trajectories over 100 iterations, extracting stable parameters ($m \approx 1.19$, $\text{DIBL} \approx 0.103$~V/V) without letting neural degrees of freedom prematurely absorb macroscopic transport trends.

\subsubsection{Stage 2: Hybrid Stochastic and Quasi-Newton Neural Training}

Following baseline localization, the baseline parameter vector $\mathbf{\Phi}$ is anchored and the neural network weight vector $\theta$ is fully unlocked. Training proceeds through two sequential sub-phases:

\begin{enumerate}
\item \textbf{Stochastic Gradient Exploration (Adam) \cite{logmse_prior}:} A stochastic gradient-based optimization algorithm with initial learning rate $\eta = 10^{-3}$ runs for 500 epochs to clear macro-structural transport features. A dynamic learning rate decay step then drops to $\eta = 10^{-4}$ for an additional 500 epochs (1000 epochs total).
\item \textbf{Deterministic Fine-Scale Localization (L-BFGS) \cite{liu1989lbfgs}:} Once Adam flattens at epoch 1000, the routine shifts to the Limited-memory Broyden--Fletcher--Goldfarb--Shanno (L-BFGS) algorithm. L-BFGS dynamically constructs a local approximation of the inverse Hessian matrix using a history of the past $m_{hist}=20$ gradient updates, eliminating residual micro-scale errors across the full temperature-bias envelope \cite{liu1989lbfgs}.
\end{enumerate}

\subsection{Evolutionary Symbolic regression and Continuum Regularization}
\label{sec:regression}

Although the trained neural network correction $\text{DNN}(\mathbf{x}; \theta)$ achieves strong numerical accuracy, deploying an active, multi-layer weight matrix directly inside a circuit simulator is computationally impractical. Evaluating hundreds of transcendental hyperbolic tangent matrix layers at every runtime step introduces unacceptable execution latency for large-scale circuit simulations.

To eliminate this performance bottleneck, we apply evolutionary symbolic regression~\cite{cranmer2023pysr} to the isolated neural correction term. During training, the unconstrained neural network handles the non-convex optimization, navigating past local minima to capture the underlying physical dynamics. Once this continuous trajectory is established, symbolic regression eliminates the multi-layer redundancy of the network, extracting a compact, analytical expression.

The genetic programming routine searches wide Pareto-optimal expression frontiers \cite{smits2005pareto} using the binary and unary operator set:
\begin{equation}
\mathcal{F} = \{+, -, \times, \div, \ln, \exp, \sqrt{\cdot}\}
\end{equation}

Rather than fixing a single linear complexity penalty -- which is highly sensitive to the loss scale of the residual and can favor trivial expressions when misspecified -- we select the deployed expression using a relative-tolerance rule applied to the Pareto frontier. Let $\text{MSE}_{\min}$ denote the lowest residual loss achieved by any candidate on the frontier (i.e., the loss of the most complex expression explored). We select the lowest-complexity candidate whose loss lies within a fixed tolerance $\tau$ of this best-achievable loss:
\begin{equation}
\mathcal{C}^\star = \min\left\{\, \mathcal{C} \;:\; \text{MSE}_{\mathcal{C}} \leq (1+\tau)\cdot \text{MSE}_{\min} \,\right\}, \qquad \tau = 0.05
\end{equation}
This criterion favors parsimony directly: additional complexity is only accepted while it still buys a non-trivial (\,$>5\%$\,) reduction in residual loss relative to the best candidate found; once loss gains fall within the noise floor of further search, the search stops adding structure. Applied to our frontier (Table~\ref{tab:pareto_frontier}), the complexity-11 expression falls outside this tolerance ($+16.3\%$ relative to the best-found loss), while the complexity-15 expression satisfies it ($+3.65\%$) and is the simplest candidate to do so, yielding the selected expression in Eq.~\ref{eq:final_symbolic_expression_formal}.

By utilizing this distilled expression (Eq.~\ref{eq:final_symbolic_expression_formal}), the final model functions as a pure analytical expression, executing at standard algebraic speeds while retaining the precise physical corrections discovered during the deep neural training phase.

\subsection{Transport Dynamics and Boundary Constraints}
\label{sec:scope}

This framework targets the automated synthesis, optimization, and numerical stabilization of continuous, non-linear steady-state and dynamic transport equations validated across a comprehensive $I$--$V$ matrix. A production-grade compact model, such as BSIM-CMG, fundamentally requires a decoupled dual architecture: a transport engine to govern channel currents ($I_{ds}$), and a charge conservation engine to govern multi-terminal displacement charges ($Q_g, Q_d, Q_s, Q_b$). The analytical derivatives of these charges define the critical device transcapacitances ($C_{gg}, C_{gd}, C_{gs}$, etc.). 

While capturing full charge conservation metrics is vital for evaluating propagation delays and dynamic power consumption in multi-stage logic architectures, modeling transport dynamics represents an independent, highly complex challenge. This complexity is driven primarily by severe subthreshold non-linearities and short-channel effects inherent to advanced nodes.

The synthesis of terminal charge equations via charge-conserving state-variable Universal Differential Equation (UDE) layers follows identical numerical principles and is reserved for separate treatment. For immediate deployment within existing circuit simulators, the synthesized symbolic transport engine proposed herein bridges directly with an unaugmented baseline PDK charge core (e.g., standard BSIM capacitance equations). This modular integration enables designers to leverage regularized $I$--$V$ transport enhancements immediately without disrupting established multi-terminal displacement charge routing configurations. Consequently, this formulation provides a self-contained solution for the automated synthesis of physically stable, simulator-convergent device transport profiles.

\section{Results and Evaluation}

We evaluate the structural accuracy, derivative continuity, and computational performance of the proposed hybrid UDE and symbolic regression workflow against a held-out evaluation partition of the MOSFET Electrical Simulation Dataset (MESD) for a 3nm multi-gate FinFET architecture.
As established in the scoping boundary (Section~\ref{sec:scope}), all reporting metrics reflect mathematical and physical fidelity with respect to the foundry-calibrated BSIM reference ground truth encoded within the MESD, providing a highly controlled and reproducible framework for automated compact model synthesis.
The evaluation focuses uniformly on N-channel configurations; P-channel compact synthesis follows identical structural paradigms but requires independent optimization tracking.

\subsection{Global Parametric Baseline Initialization}

During Stage 1 optimization, the physical parameter vector $\mathbf{\Phi}$ governing the macroscopic electrostatic baseline was extracted by executing a PSO global search over the complete MESD multi-bias training matrix with the neural network correction locked to zero ($\theta = \mathbf{0}$). This isolates the fundamental continuum-level boundary physics before introducing data-driven structural corrections. The primary extracted parameters are summarized in Table~\ref{tab:extracted_parameters}.

\begin{table}[H]
\centering
\caption{Extracted Primary Physical Baseline Parameters (Stage 1 PSO)}
\label{tab:extracted_parameters}
\begin{tabular}{lllc}
\toprule
\textbf{Parameter} & \textbf{Physical Description} & \textbf{Extracted Value} & \textbf{Units} \\
\midrule
$\mu_{00}$ & Baseline Low-Field Intrinsic Electron Mobility & $0.01000$ & $\text{m}^2/\text{V}\cdot\text{s}$ \\
$V_{th0}$ & Intrinsic Zero-Bias Flatband Threshold Voltage & $0.18859$ & V \\
$m$ & Subthreshold Ideality Factor (Gate Control Swing) & $1.19241$ & $-$ \\
$\text{DIBL}$ & Drain-Induced Barrier Lowering Coefficient & $0.10314$ & $\text{V}/\text{V}$ \\
$C_{ox}$ & Effective Gate Dielectric Oxide Capacitance & $0.03365$ & $\text{F}/\text{m}^2$ \\
\bottomrule
\end{tabular}
\end{table}

The extracted low-field mobility $\mu_{00} = 0.01$~m$^2$/V$\cdot$s ($100$~cm$^2$/V$\cdot$s) aligns with electron transport limits in highly constrained multi-gate silicon geometries at room temperature \cite{jacoboni1977review}. The baseline zero-bias threshold voltage $V_{th0} = 0.18859$~V and subthreshold ideality factor $m = 1.19241$ are physically consistent with a high-performance 3nm NMOS FinFET. We note that the MESD \texttt{N3A} partition sweeps terminal biases only up to $0.7$~V, so all extracted parameters are anchored to that envelope; no claim is made about a $1.2$~V rail. The effective gate oxide capacitance $C_{ox} = 0.03365$~F/m$^2$ corresponds to an Equivalent Oxide Thickness (EOT) of approximately $1.02$~nm, matching standard industry metrics for advanced high-$\kappa$/metal-gate (HKMG) stacks \cite{chau2004highk}.

By anchoring these macroscopic electrostatic parameters during Stage 1, the subsequent neural training phase is mathematically constrained; the machine learning capacities are restricted from absorbing known boundary behaviors, forcing the neural field to isolate and extract only the high-dimensional transport anomalies that legacy closed-form equations fail to resolve.

\subsection{Global Calibration Performance}
\label{sec:results_accuracy}

Following the immobilization of the macroscopic baseline parameters, the neural
correction network was unlocked and optimized using the multi-decade $\log_{10}$-space
loss function of Eq.~\eqref{eq:log_loss}. Table~\ref{tab:global_performance} provides a rigorous accuracy comparison across three distinct modeling configurations: the isolated Caughey--Thomas (CT) baseline, the full hybrid UDE containing the active neural network field, and the compiled Symbolic Model where the deep neural network is replaced entirely by the distilled closed-form expression $\Delta\mu_{\text{extracted}}$. Errors are computed as
\begin{equation}
\text{MRE} = \frac{1}{N}\sum_{i=1}^{N} \frac{\left|I_{\text{sim},i} - I_{\text{MESD},i}\right|}{|I_{\text{MESD},i}| + \epsilon}, \qquad
\text{MSLE} = \frac{1}{N}\sum_{i=1}^{N}\left[\log_{10}(I_{\text{sim},i}+\epsilon) - \log_{10}(|I_{\text{MESD},i}|+\epsilon)\right]^2
\end{equation}
where $\epsilon = 10^{-15}$~A is a small floor added to prevent division by, or logarithms of, near-zero currents; this choice disproportionately affects the deep off-state regime and should be reported explicitly.

\begin{table}[H]
\centering
\caption{Global Model Performance Metrics across Optimization Boundaries}
\label{tab:global_performance}
\begin{tabular}{lcccc}
\toprule
& \multicolumn{2}{c}{\textbf{MESD Training Partition}} & \multicolumn{2}{c}{\textbf{Held-Out Evaluation Partition}} \\
\cmidrule(lr){2-3} \cmidrule(lr){4-5}
\textbf{Model Configuration} & \textbf{RMSE (A)} & \textbf{MRE (\%)} & \textbf{RMSE (A)} & \textbf{MRE (\%)} \\
\midrule
CT Baseline Scaffolding (Stage 1) & $8.44 \times 10^{-5}$ & $73.82\%$ & $6.94 \times 10^{-5}$ & $70.33\%$ \\
Neural Hybrid UDE (Stage 2)       & $1.93 \times 10^{-5}$ & $10.45\%$ & $1.73 \times 10^{-5}$ & $11.01\%$ \\
\textbf{Symbolic Model (Deployed SPICE Expression)} & $\mathbf{2.07 \times 10^{-5}}$ & $\mathbf{11.83\%}$ & $\mathbf{1.91 \times 10^{-5}}$ & $\mathbf{12.38\%}$ \\
\bottomrule
\end{tabular}
\end{table}

The deployment of the neural UDE drives a substantial reduction in the held-out evaluation Mean Relative Error (MRE), collapsing from $70.33\%$ down to $11.01\%$. The synthesized Symbolic Model, which constitutes the distilled closed-form algebraic transport expression evaluated inside SPICE, achieves a held-out test MRE of $12.38\%$, representing a negligible accuracy degradation of only $1.37$ percentage points relative to the neural baseline.

The tight tracking performance observed between the training and test partitions ($<0.6\%$ MRE divergence for the symbolic model) demonstrates that the hybrid framework captures genuine, systematic nanoscale transport physics rather than overfitting localized statistical fluctuations.

\begin{figure}[H]
\centering
\begin{tabular}{cc}
\includegraphics[width=0.48\textwidth]{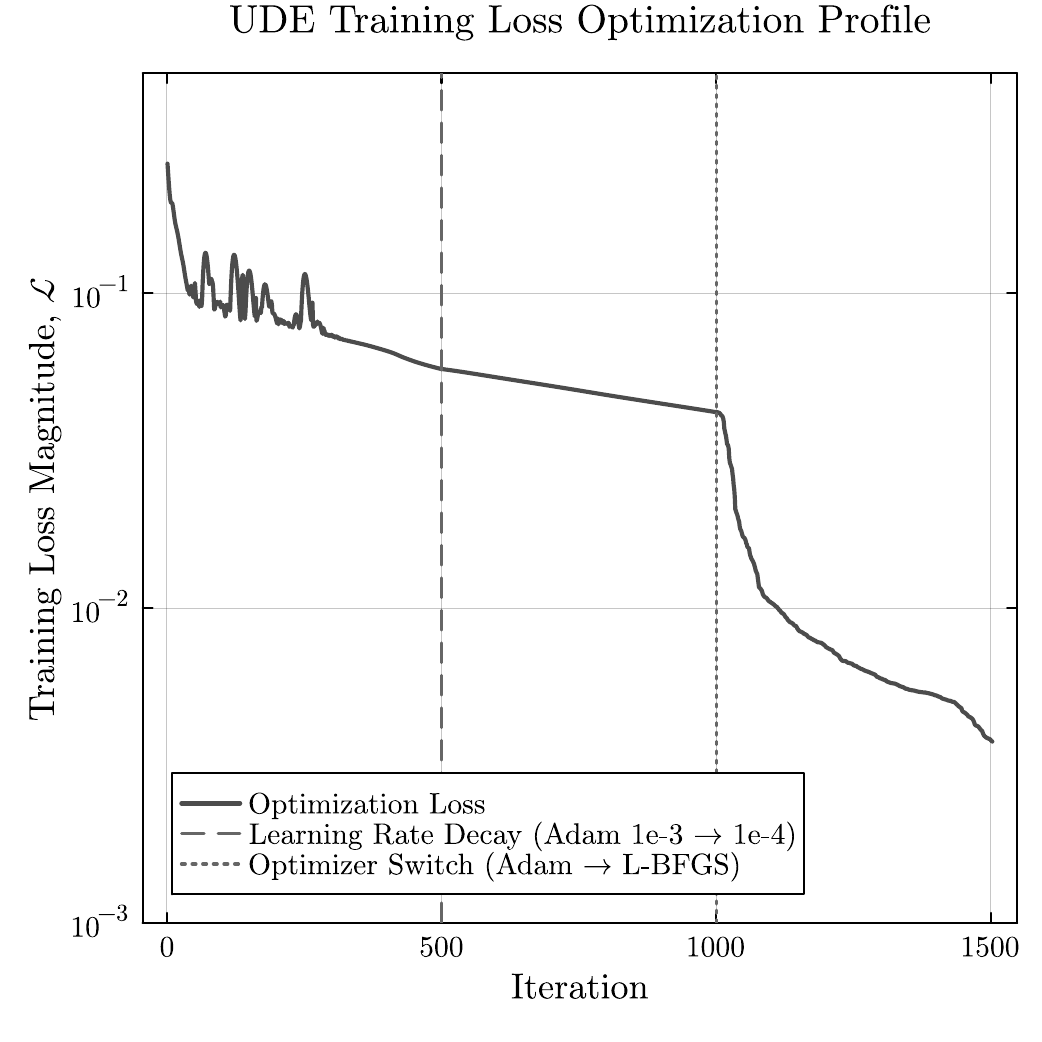} &
\includegraphics[width=0.48\textwidth]{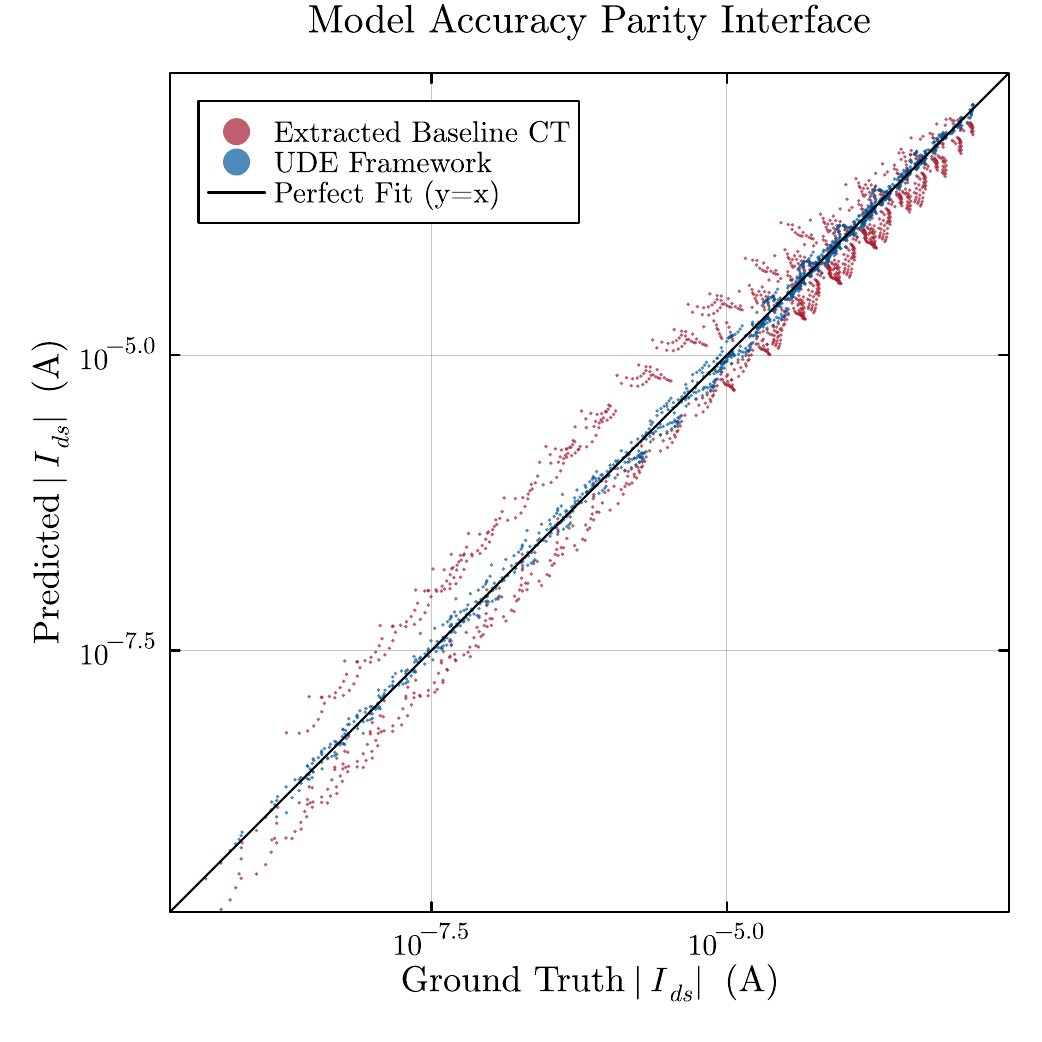} \\
(a) & (b)
\end{tabular}
\caption{Multi-decade training loss descent (a) and global prediction parity (b) for the neural UDE optimization. Panel (a) tracks the log-space loss $\mathcal{L}_{\text{log}}$ over training epochs. Panel (b) compares predicted versus reference $I_{ds}$ across the full training sweep, with the diagonal indicating good agreement.}
\label{fig:training_loss}
\end{figure}

The initial $70.33\%$ error profile of the unaugmented baseline reflects the intentional simplification of the Caughey--Thomas scaffold. The purpose of this architecture is not to independently match multi-parameter standards like BSIM-CMG, but to establish a computationally stable mathematical skeleton that anchors boundary states and prevents matrix divergence, while the UDE pipeline automates parameter calibration by learning the complete residual physics. The multi-decade training loss descent and the resulting global prediction parity are illustrated in Fig.~\ref{fig:training_loss}.

\subsection{Multi-Bias Transfer Characteristics}

\begin{figure}[H]
\centering
\includegraphics[width=\textwidth]{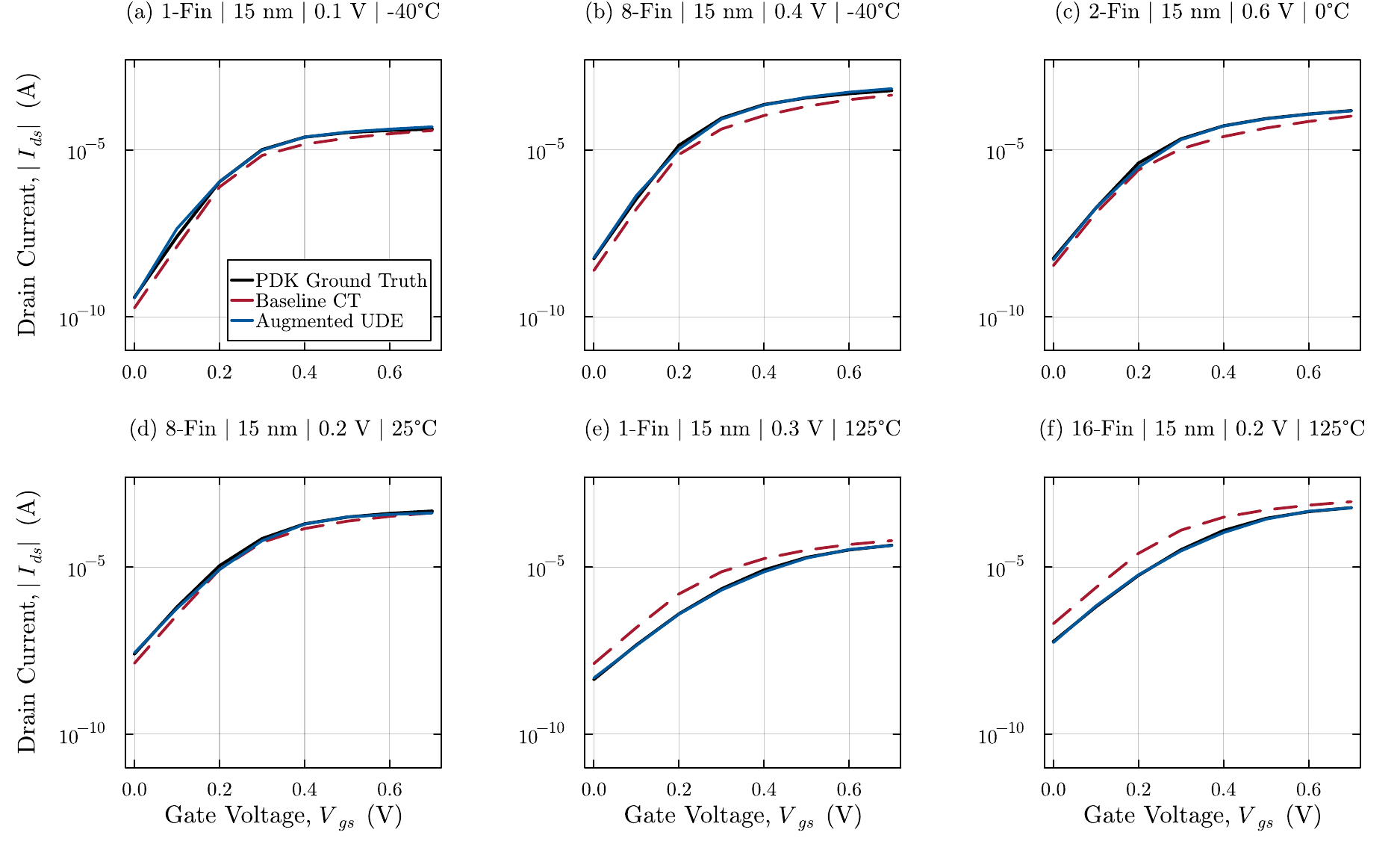}
\caption{Multi-panel validation matrix profiling $I_{ds}$--$V_{gs}$ transfer characteristics across independent drain-to-source biases ($V_{ds} = 0.1$~V, $0.2$~V, $0.3$~V, $0.4$~V, $0.6$~V) and thermal operational boundaries ($T = -40^\circ\text{C}, 0^\circ\text{C}, 25^\circ\text{C}, 125^\circ\text{C}$). Solid blue lines track the neural UDE model prior to symbolic regression, solid black lines represent the MESD reference ground truth, and dashed red lines mark the unaugmented physical baseline scaffolding. The tight alignment between the \emph{neural UDE} and the reference data across all coordinates demonstrates accurate multi-variable interpolation and physical scaling without artifacts at the operational boundaries of the MESD envelope. The corresponding accuracy of the distilled symbolic expression is reported separately in Table~\ref{tab:global_performance} (configuration \textit{A8}).}
\label{fig:main_iv_curves_composite}
\end{figure}

As demonstrated in the comprehensive transfer tracking matrix of Fig.~\ref{fig:main_iv_curves_composite}, the unaugmented Caughey--Thomas baseline suffers from severe, systematic structural deviations across the entire multi-decade bias landscape, underestimating drive current in strong inversion and distorting the subthreshold slope profile. Conversely, the neural UDE model replicates the reference ground truth characteristics across all plotted operational domains. The model matches the exponential subthreshold leakage regime, handles the moderate-inversion curvature, and captures the velocity-saturated drive current at the highest drain bias present in the MESD partition ($V_{ds} = 0.6$~V). This tracking fidelity demonstrates the framework's ability to interpolate accurately across combinations of bias and temperature not encountered during optimization; behaviour beyond the $0.7$~V MESD bias ceiling is extrapolation and is examined separately in Section~\ref{sec:spice_impl}.

\subsection{Error Distribution Across Device Operating Conditions}

To verify the mathematical uniformity of the optimization landscape, Table~\ref{tab:ude_breakdown} tracks the Mean Squared Logarithmic Error (MSLE) across independent structural and environmental boundary slices.

\begin{table}[H]
\centering
\caption{Detailed Log-Domain Tracking Error Breakdown ($\text{MSLE}_{\text{UDE}}$)}
\label{tab:ude_breakdown}
\begin{tabular}{lcc}
\toprule
\textbf{Operational Evaluation Slice} & \textbf{Structural Parameter Condition} & \textbf{MSLE} \\
\midrule
Extreme Geometric Gate Limit & $N_{fin} = 16$ & $3.7637 \times 10^{-3}$ \\
Nominal Room Thermal State   & $T = 25^\circ\text{C}$  & $4.5794 \times 10^{-3}$ \\
Extended Thermal Boundary    & $T = 125^\circ\text{C}$ & $5.1208 \times 10^{-3}$ \\
\bottomrule
\end{tabular}
\end{table}

The tracking error remains uniformly bounded across all structural configurations, with MSLE values consistently clustering near the $10^{-3}$ decade. The absence of localized error spikes confirms that the internal neural network discovered a balanced physical correction vector that regularizes effectively across both geometric scale and thermal boundaries.

\subsection{Thermal Generalization Analysis}

\begin{figure}[H]
\centering
\includegraphics[width=0.8\textwidth]{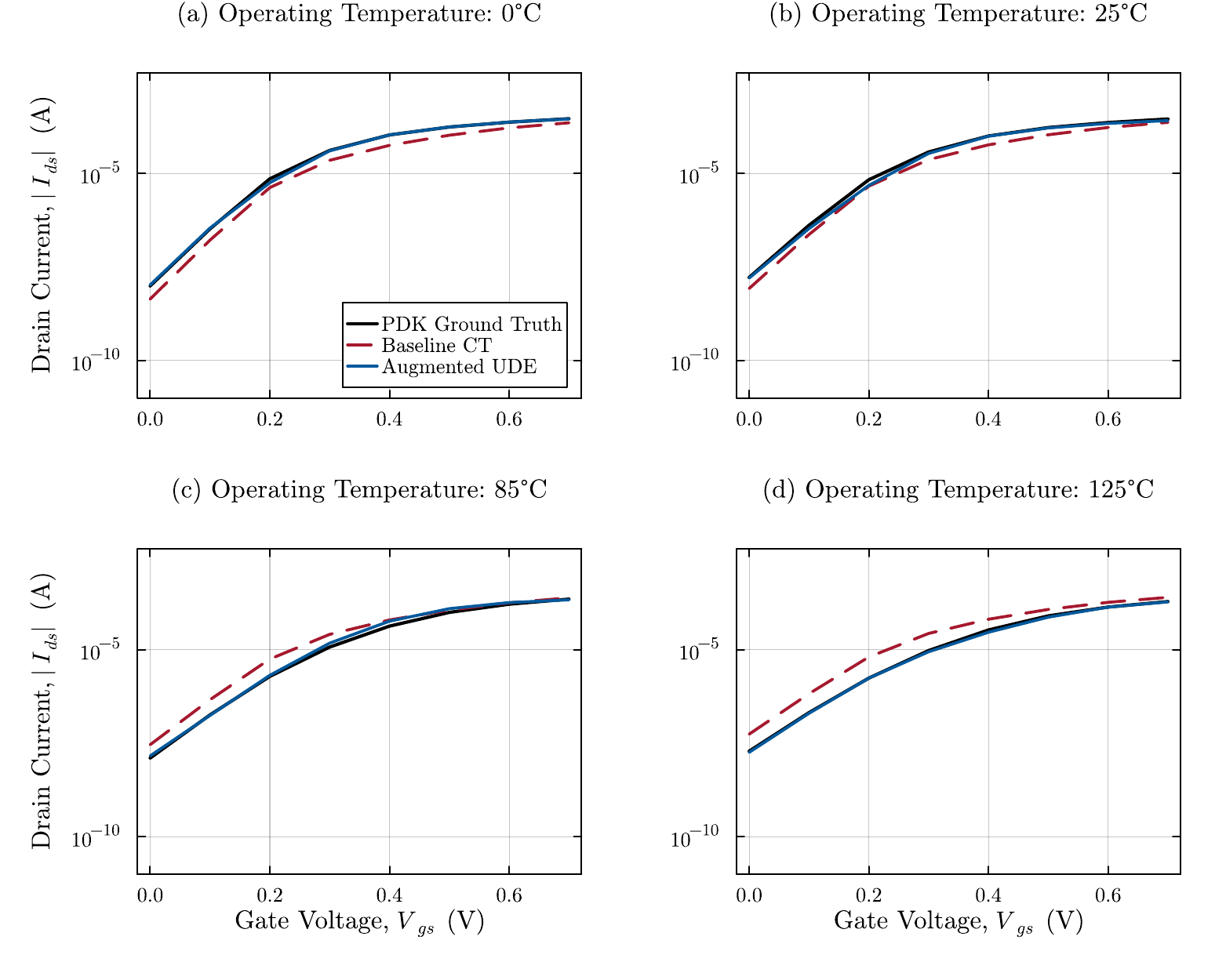}
\caption{Thermal generalization sweeps profiling subthreshold performance and threshold voltage shifting across a operating envelope ($0^\circ\text{C}$ to $125^\circ\text{C}$) at: (a) $0^\circ\text{C}$, (b) $25^\circ\text{C}$, (c) $85^\circ\text{C}$, and (d) $125^\circ\text{C}$. Solid blue lines track our neural UDE model prior to symbolic regression, solid black lines represent the reference ground truth, and dashed red lines mark the unaugmented Caughey--Thomas baseline. The plots demonstrate that the physics-guided structure accurately captures subthreshold swing degradation and weak-inversion leakage across all thermal states where the baseline fails.}
\label{fig:generalization_heatmap}
\end{figure}

The automated compact framework preserves temperature-dependent current trajectories despite complex physical changes in underlying device parameters, such as acoustic phonon mobility degradation \cite{jacoboni1977review}, thermal velocity saturation limits \cite{caughey1967mobility}, and bandgap contraction \cite{tsividis1999mosfet}. This thermal fidelity is visualized across the full environmental range in Fig.~\ref{fig:generalization_heatmap}.
Because the temperature scaling parameter $\text{TRatio}$ is embedded directly into both
the physical baseline equations and the input vector of the neural network, the neural
UDE plotted in Fig.~\ref{fig:generalization_heatmap} tracks continuous threshold shifting
and leakage variation across the environmental range; the distilled symbolic expression
inherits this behaviour with the accuracy penalty quantified in
Table~\ref{tab:global_performance}.

\subsection{Subthreshold Swing}

The preservation of the subthreshold swing (SS) is a critical standard for circuit-level compact modeling, defining the gate control efficiency and off-state power boundaries in weak inversion:
\begin{equation}
\text{SS} = \left(\frac{d\log_{10}(I_{ds})}{dV_{gs}}\right)^{-1} \quad [\text{mV/dec}]
\end{equation}
Table~\ref{tab:ss_comparison} evaluates the statistical distribution of the subthreshold swing across all evaluated geometric configurations, confirming that the model maintains high derivative fidelity.

\begin{table}[H]
\centering
\caption{Statistical Distribution of Subthreshold Swing (SS)}
\label{tab:ss_comparison}
\begin{tabular}{lcc}
\toprule
\textbf{Modeling Framework} & \textbf{Mean SS $\pm$ Std.\ Dev.\ (mV/dec)} & \textbf{Absolute Mean Discrepancy} \\
\midrule
MESD Reference Ground Truth  & $79.6 \pm 15.5$ & \textit{Base Standard} \\
Extracted Baseline Scaffolding & $76.9 \pm 14.3$ & $-2.7$~mV/dec \\
\textbf{Synthesized UDE Symbolic Model (Ours)} & $\mathbf{81.0 \pm 14.4}$ & $\mathbf{+1.4}$~\textbf{mV/dec} \\
\bottomrule
\end{tabular}
\end{table}

The compiled symbolic compact model accurately reproduces the mean and standard deviation of the reference distribution, achieving an absolute mean discrepancy of just $+1.4$~mV/dec. This demonstrates that the UDE optimization pipeline preserves essential first-order derivative trajectories ($\partial I_{ds} / \partial V_{gs}$) across multi-decade transitions, avoiding the localized transconductance discontinuities that often cause convergence failures in purely data-driven models.

\subsection{Complete Structural Ablation Matrix and Architectural Sensitivity}

To evaluate the mathematical necessity and physical consistency of each component within the proposed physics-augmented framework, a comprehensive structural ablation study was conducted. The model's performance was evaluated across ten distinct configurations divided into four primary investigative vectors: physical-only baselines, unconstrained deep learning surrogates, hyperparameter operator width sweeps, and robustness boundaries (training-set reduction and thermal-extreme evaluation). 

To ensure rigorous physical compliance, each configuration was evaluated using three performance metrics: Root Mean Squared Error (RMSE) to quantify absolute on-state matching, Mean Relative Error (MRE) to evaluate uniform tracking across all seven decades of operational current, and the Max Zero-Bias Ghost Current ($I_{\text{ghost}}$) to monitor boundary condition preservation when $V_{ds} \equiv 0$~V. The complete performance matrix is summarized in Table~\ref{tab:complete_ablation_matrix}.

\begin{table}[H]
\centering
\caption{Complete Structural Ablation and Architectural Sensitivity Performance Matrix (Held-Out Evaluation Partition)}
\label{tab:complete_ablation_matrix}
\resizebox{\textwidth}{!}{%
\begin{tabular}{lllccr}
\toprule
\textbf{ID} & \textbf{Configuration Variant} & \textbf{Core Architectural/Data Modification} & \textbf{RMSE (A)} & \textbf{MRE (\%)} & \textbf{Max $I_{\text{ghost}}$ at $V_{\text{ds}}=0$~V (A)} \\
\midrule
\textit{A1} & Physical Baseline & Baseline Caughey--Thomas drift-diffusion scaffolding only & $6.94 \times 10^{-5}$ & $70.33\%$ & $0.00$ (Preserved) \\
\textit{A2} & Pure Black-Box MLP & Unconstrained 2-hidden layer neural network surrogate & $5.54 \times 10^{-3}$ & $2.06 \times 10^{7}\%$ & $3.86 \times 10^{-2}$ (\textbf{VIOLATED}) \\
\textit{A3} & Linear Loss Ablation & Identical UDE model optimized via standard linear MSE loss & $6.40 \times 10^{-5}$ & $159.91\%$ & $0.00$ (Preserved) \\
\textit{A4} & \textbf{PRISM-UDE (Ours)} & \textbf{Continuous neural transport operator embedded in scaffold (64 hidden neurons)} & $\mathbf{1.73 \times 10^{-5}}$ & $\mathbf{11.01\%}$ & $\mathbf{0.00}$ \textbf{(Preserved)} \\
\midrule
\multicolumn{6}{l}{\textit{Compression, Training-Set Reduction, and Thermal-Extreme Evaluation}} \\
\textit{A5} & Hidden Width = 8  & Reduced capacity neural operator (8 hidden neurons)   & $2.60 \times 10^{-5}$ & $14.11\%$ & $0.00$ (Preserved) \\
\textit{A6} & Hidden Width = 16 & Moderate capacity neural operator (16 hidden neurons)  & $2.16 \times 10^{-5}$ & $11.70\%$ & $0.00$ (Preserved) \\
\textit{A7} & Hidden Width = 32 & Intermediate capacity neural operator (32 hidden neurons) & $2.04 \times 10^{-5}$ & $11.12\%$ & $0.00$ (Preserved) \\
\midrule
\multicolumn{6}{l}{\textit{Compression, Data Efficiency, and Extreme Domain Extrapolation}} \\
\textit{A8} & Symbolic regression & Closed-form Complexity-15 analytical Pareto compact expression & $1.91 \times 10^{-5}$ & $12.38\%$ & $0.00$ (Preserved) \\
\textit{A9} & 50\% Training-Set Reduction & Retrained on half of the 140 training trajectories (70 trajectories) & $1.89 \times 10^{-5}$ & $10.40\%$ & $0.00$ (Preserved) \\
\textit{A10} & Extreme-Temperature Evaluation & Held-out evaluation restricted to the thermal extremes ($0^{\circ}\text{C}$, $125^{\circ}\text{C}$) & $9.67 \times 10^{-6}$ & $6.57\%$ & $0.00$ (Preserved) \\
\bottomrule
\end{tabular}
}
\end{table}

\subsubsection{Comparison of Baseline, Unconstrained Surrogates, and PRISM-UDE}

A critical insight from the ablation study is the failure of unconstrained deep learning architectures in semiconductor device modeling. The pure black-box multilayer perceptron (\textit{A2}) yields an unacceptably high mean relative error exceeding $2.06 \times 10^{7}\%$. This mathematical collapse occurs because a conventional neural network minimized on an unweighted landscape treats the deep subthreshold and off-state regimes ($10^{-12}$ to $10^{-9}$~A) as numerical zero, failing to capture the exponential physics of subthreshold swing. 

More importantly, \textit{A2} severely violates fundamental boundary laws, generating a massive $38.6$~mA ghost current at $V_{\text{ds}} = 0$~V. In contrast, the baseline physical scaffolding (\textit{A1}) guarantees a mathematically clean zero-current state ($0.00$~A), but suffers from an elevated MRE ($70.33\%$) due to the inability of classical Caughey--Thomas equations to capture nanoscale source-drain coupling and transport degradation. By embedding the neural operator directly inside the multiplicative drift-diffusion boundary layer template, the full neural PRISM-UDE (\textit{A4}) corrects the physical baseline's structural error, slashing the relative error to $11.01\%$, while preserving the zero-bias constraint.

\subsubsection{The Effect of Logarithmic Loss Formulation}

Evaluating the model under a standard linear MSE objective (\textit{A3}) demonstrates that optimization formulation is as crucial as architectural design. Training an identical UDE model with a linear loss function degrades the MRE to $159.91\%$. Because linear MSE scales strictly with absolute current magnitudes, the optimization gradients are driven almost entirely by the high-current strong-inversion saturation domain ($10^{-3}$~A). 

Consequently, multi-decade discrepancies in the subthreshold leakage domain are treated as negligible noise. Mapping the error space uniformly through log-domain regularization balances gradient pressure evenly across all seven decades of device operation, providing a $14.5\times$ improvement in relative precision ($159.91\% \to 11.01\%$).

\subsubsection{Effect of Hidden-Layer Width}

Sweeping the neural network's hidden layer width across configurations \textit{A5}, \textit{A6}, \textit{A7}, and \textit{A4} (Width = 64) reveals a highly stable, monotonic convergence curve. As the hidden unit dimensional capacity expands from $8 \to 16 \to 32 \to 64$, the relative error tracks smoothly through $14.11\% \to 11.70\% \to 11.12\% \to 11.01\%$. This asymptotic trend demonstrates that the underlying physics-informed architecture is highly robust against over-parameterization. The accuracy improvement between 32 and 64 hidden units is modest but consistent, and the choice of 64 hidden neurons (A4) as the deployed configuration reflects the favourable accuracy-to-complexity trade-off at this operator width.

\subsubsection{Compression, Data Efficiency, and Resilience}

The final analytical vector demonstrates the framework's exceptional structural stability under constraints of model compression, data scarcity, and environmental extrapolation:
\begin{itemize}
    \item \textbf{Symbolic Compression (\textit{A8}):} Compressing the continuous neural field into the final distilled, closed-form algebraic compact expression via symbolic regression incurs a negligible relative accuracy penalty, with the MRE shifting slightly from $11.01\%$ to $12.38\%$. This minimum information delta proves that the synthesized expression accurately extracts the true physical transport corrections from the neural network weights without sacrificing interpretability or execution speed.
    \item \textbf{Training-Set Reduction (\textit{A9}):} Retraining on half of the training trajectories (70 of 140) yields an MRE of $10.40\%$, statistically indistinguishable from the full-data result of $11.01\%$. Halving the trajectory count leaves accuracy unchanged because the drift-diffusion scaffold already supplies the macroscopic bias and temperature trends, so the residual the network must learn is low-dimensional. We report this as insensitivity to a $2\times$ reduction in training trajectories; the experiment does not establish the sample-complexity floor of the method, and no comparison against an unconstrained surrogate under matched data budgets was run.
    \item \textbf{Extreme-Temperature Evaluation (\textit{A10}):} Restricting the held-out evaluation to the thermal extremes of the sweep ($0^{\circ}\text{C}$ and $125^{\circ}\text{C}$) yields an MRE of $6.57\%$. Because both temperatures are represented in the training partition, this is an evaluation at the edges of the sampled thermal envelope rather than out-of-distribution extrapolation. It shows that accuracy does not degrade at the corners of the temperature grid, which is the behaviour the embedded $\text{TRatio}$ dependence in the physical baseline is intended to enforce.
\end{itemize}

\section{Symbolic regression and Compact Model Recovery}

\subsection{Pareto Evolutionary Optimization Space}

\begin{figure}[H]
\centering
\includegraphics[width=0.72\textwidth]{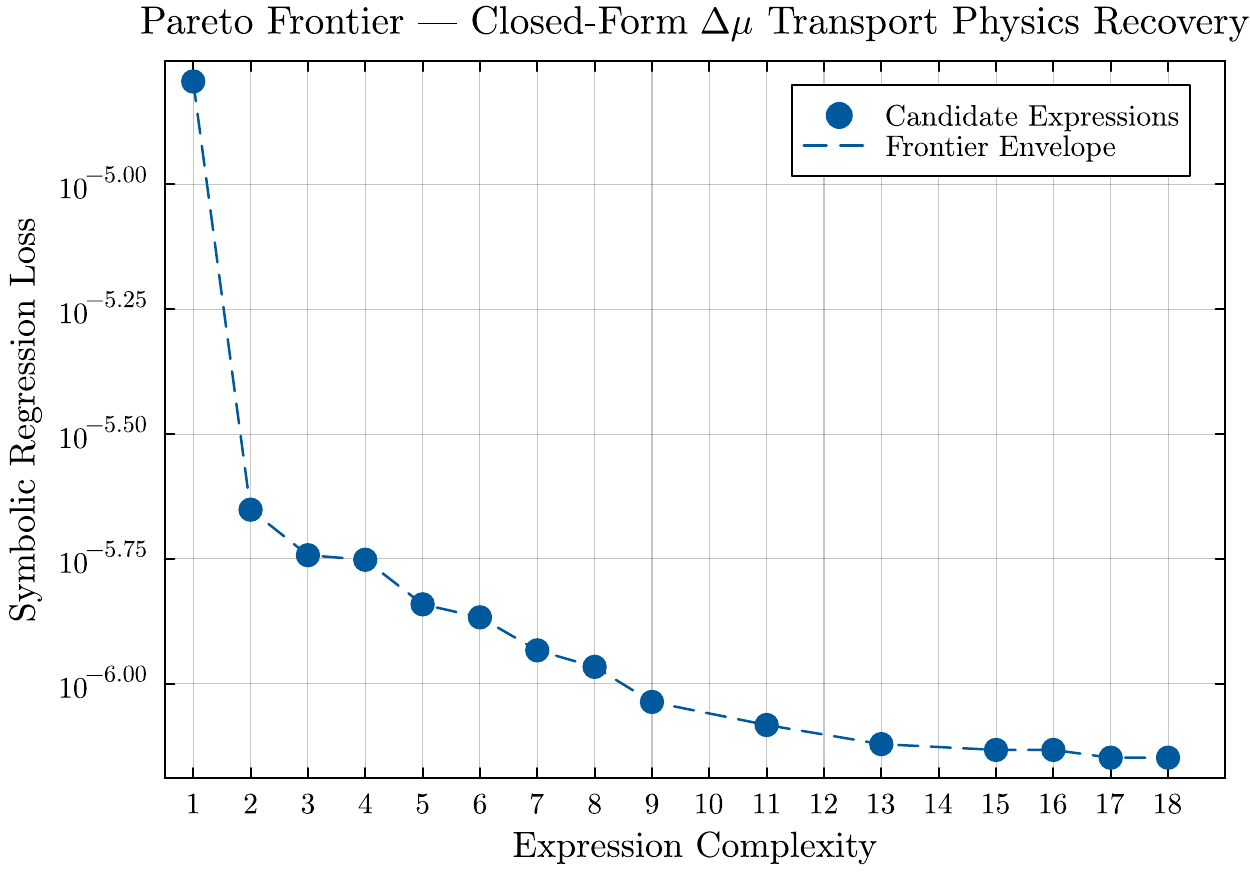}
\caption{Model selection and compression trade-off verified via the Pareto regression frontier. The chart maps algebraic complexity against symbolic regression residual loss, highlighting the selected optimal expression at a complexity score of 15. The frontier illustrates that the framework compresses the neural network into a compact functional form with a negligible accuracy penalty.}
\label{fig:pareto_frontier}
\end{figure}

We applied Pareto-optimal symbolic regression directly to the continuous neural field $\text{DNN}(\mathbf{x}; \theta)$ to extract a clean mathematical expression for the residual transport correction $\Delta\mu_{\text{extracted}}$. The evolutionary search balanced algebraic complexity $\mathcal{C}$ against tracking loss to identify the optimal expression layout. Table~\ref{tab:pareto_frontier} presents candidate expressions discovered along the Pareto frontier, and the complexity--loss trade-off is plotted in Fig.~\ref{fig:pareto_frontier}.

\begin{table}[H]
\centering
\caption{Pareto Frontier for Extracted Mobility Field Correction}
\label{tab:pareto_frontier}
\small
\begin{tabular}{ccl}
\toprule
\textbf{Complexity ($\mathcal{C}$)} & \textbf{Residual Loss} & \textbf{Recovered Closed-Form Expression} \\
\midrule
1  & $1.607 \times 10^{-5}$ & $y = \mu_{CT}$ \\
2  & $2.231 \times 10^{-6}$ & $y = \ln(0.99997)$ \\
3  & $1.808 \times 10^{-6}$ & $y = -0.20810 \cdot \mu_{CT}$ \\
5  & $1.442 \times 10^{-6}$ & $y = \mu_{CT} \cdot (0.96342 - \text{TRatio})$ \\
6  & $1.358 \times 10^{-6}$ & $y = (V_{ds} - \ln(\text{TRatio})) \cdot \mu_{CT}$ \\
9  & $9.192 \times 10^{-7}$ & $y = (V_{ds} - (\text{TRatio} - 0.78803)) \cdot (\mu_{CT}/\text{TRatio})$ \\
11 & $8.265 \times 10^{-7}$ & $y = (\mu_{CT}/\text{TRatio}) \cdot [(V_{ds}/\text{TRatio}) - (\text{TRatio} - 0.78489)]$ \\
\textbf{15} & $\mathbf{7.367 \times 10^{-7}}$ & $\mathbf{y = \sqrt{\mu_{CT}^2 + 10^{-6}} \cdot \left[\dfrac{V_{ds}/1.2}{\text{TRatio}^2} - \ln\!\left(\text{TRatio} + \dfrac{0.06499}{V_{gs}/1.2 + 0.07219}\right)\right]}$ \\
17 & $7.108 \times 10^{-7}$ & $y = \mu_{CT} \cdot [(V_{ds}/\text{TRatio}) + (1.4944 V_{ds} - \text{TRatio}^2) \cdot (V_{ds}/\text{TRatio})]$ \\
\bottomrule
\end{tabular}
\end{table}

\subsection{Optimal Compact Expression and Physical Interpretation}

To establish a rigorous physical framework for the discovered transport correction, the empirical expression at \textbf{Complexity 15} is cast into a generalized analytical form. By isolating the numerical coefficients into a set of dedicated physical parameter constants ($\alpha_i$), we decouple the core transport mechanics from the specific technology normalization factors. 

The finalized, extracted transport correction formula is defined as Eq.~\ref{eq:generalized_transport_correction}, decomposing the correction into a mobility scaling pre-factor and two competing field-dependent terms:
\begin{equation}
\Delta\mu_{\text{extracted}} = \mathcal{M}(\mu_{CT}) \cdot \left[ \mathcal{F}_{\text{drift}}(V_{ds}, T) - \mathcal{F}_{\text{gate}}(V_{gs}, T) \right]
\label{eq:generalized_transport_correction}
\end{equation}

Where the constituent scaling functions are explicitly written as:
\begin{equation}
\Delta\mu_{\text{extracted}} = \sqrt{\mu_{CT}^2 + \alpha_0} \cdot \left[\frac{V_{ds} / V_{dd}}{\alpha_1 \cdot \theta_T^2} - \ln\!\left(\theta_T + \frac{\alpha_2}{V_{gs}/V_{dd} + \alpha_3}\right)\right]
\label{eq:final_symbolic_expression_formal}
\end{equation}
Here, $\mu_{CT} \equiv \mu_0(V_{ds}, \text{TRatio})$ represents the baseline Caughey--Thomas low-field/high-field mobility defined in Eq.~(\ref{eq:ct}). $V_{dd} = 1.2\text{ V}$ is the fixed bias normalization constant carried over from the network input scaling (Section~\ref{sec:ude_theory}) rather than an operating supply rail, and $\theta_T = \text{TRatio} = T/T_0$ defines the normalized temperature ratio relative to the reference thermodynamic temperature $T_0 = 300\text{ K}$. The regularizer $\alpha_0$ carries units of $\text{m}^{4}\text{V}^{-2}\text{s}^{-2}$, matching $\mu_{CT}^{2}$. All mobilities in Eq.~\eqref{eq:final_symbolic_expression_formal} are expressed in SI units ($\text{m}^{2}\text{V}^{-1}\text{s}^{-1}$), and the numerical value $\alpha_0=10^{-6}$ is quoted on that basis in Table~\ref{tab:model_constants} and used unchanged in the SPICE implementation of Section~\ref{sec:spice_impl}. The deployed effective mobility evaluated inside SPICE is explicitly constructed as $\mu_{\text{eff, deployed}} = \mu_0(V_{ds}, \text{TRatio}) + \Delta\mu_{\text{extracted}}$.

Mathematically, the inclusion of the inverse-quadratic temperature scaling ($\theta_T^{-2}$) coupled to the drain field ($V_{ds}$) uncovers a clear structural alignment with physical lattice scattering mechanisms and severe high-field velocity saturation anomalies characteristic of short-channel 3nm multi-gate architectures. Similarly, the logarithmic sub-expression provides a physically motivated structure for vertical gate-field screening trends, capturing the severe degradation of carrier mobility under strong inversion thresholds without resorting to high-order polynomial expansions that risk non-monotonic divergence. While these functional forms align well with known physical phenomena—such as gate-field screening, subband splitting, or ballistic transport effects—they represent candidate physical interpretations consistent with the operational empirical data rather than directly observed microscopic mechanisms. A detailed discussion of these physical interpretations and their theoretical alignment is provided in Section~\ref{sec:physical_insights}.

\subsection{Extraction and Physics of Model Constants}
The extracted optimal values for the regularizing and dimensionless parameter constants are summarized in Table~\ref{tab:model_constants}, establishing the formal mapping between the mathematical fit and its physical boundaries.

\begin{table}[htbp]
\caption{Extracted Constants and Physical Governance for the Discovered Transport Model}
\label{tab:model_constants}
\centering
\begin{tabular}{lccl}
\hline
\textbf{Constant} & \textbf{Extracted Value} & \textbf{Dimension} & \textbf{Physical Significance \& Governance} \\ \hline
$\alpha_0$ & $10^{-6}$ & $\text{m}^{4}\cdot\text{V}^{-2}\cdot\text{s}^{-2}$ & Smoothness regularizer for $\mathcal{C}^\infty$ continuity at zero-bias conditions. \\
$\alpha_1$ & $1.0000$  & Dimensionless & High-field horizontal drift coupling coefficient. \\
$\alpha_2$ & $0.06499$ & Dimensionless & Quantum subband splitting and vertical field degradation parameter. \\
$\alpha_3$ & $0.07219$ & Dimensionless & Low-bias gate-field screening saturation parameter. \\ \hline
\end{tabular}
\end{table}

Crucially, while the functional \textit{form} discovered by the genetic programming loop uncovers elegant, physically interpretable scaling paths (such as the non-linear thermal suppression of mobility), a vital engineering caveat must be applied to these extracted parameters. These numeric scalar constants are inherently empirical fits tightly bound to the nominal parameter target space and structural boundaries of the anonymized ``N3A'' baseline process node within the MESD dataset. They do not represent universal physical invariants; rather, they demonstrate the successful capacity of the UDE framework to automatically distill highly localized, geometry-dependent nanoscale transport anomalies into predictable, simulator-stable algebraic parameters.

\section{Analysis of Discovered Expressions}
\label{sec:physical_insights}

This distilled expression provides deep, non-trivial physical insights into the machine-learned transport corrections, mapping directly to advanced semiconductor transport theory:

\begin{enumerate}
    \item \textbf{Continuous Regularized Mobility Scaling:} The pre-factor $\mathcal{M}(\mu_{CT}) = \sqrt{\mu_{CT}^2 + \alpha_0}$ functions as a mathematically smooth absolute operator. This forces the correction factor to scale symmetrically and monotonically with the magnitude of the baseline Caughey--Thomas mobility \cite{caughey1967mobility}. The regularization constant $\alpha_0 = 10^{-6}~\text{m}^4\text{V}^{-2}\text{s}^{-2}$ is four decades below $\mu_{00}^2$ at the extracted baseline mobility, making it inert wherever mobility is appreciable while guaranteeing infinite differentiability ($\mathcal{C}^\infty$ continuity) across zero-bias and velocity-reversal states \cite{nocedal2006numerical}. This eliminates numerical derivative discontinuities, ensuring robust convergence during SPICE circuit simulations.
    
    \item \textbf{Non-Local Drain-Field--Thermal Cross-Coupling:} The term $\mathcal{F}_{\text{drift}} = V_{ds}/(\alpha_1 \cdot V_{dd} \cdot \theta_T^2)$ is consistent with non-local transport corrections under high horizontal electric fields. The inverse-quadratic temperature dependence ($\theta_T^{-2}$) is compatible with suppression of non-local velocity-overshoot or ballistic contributions at elevated temperatures, mirroring the qualitative trend expected from enhanced acoustic and optical phonon scattering \cite{jacoboni1977review}. We note this is one plausible interpretation of the fitted functional form rather than a mechanism uniquely identified by the regression, which was fit to terminal $I$--$V$ data alone.
    
    \item \textbf{Gate-Controlled Logarithmic Carrier Modulation:} The logarithmic transfer function, 
    \begin{equation}
    \mathcal{F}_{\text{gate}} = \ln\!\left(\theta_T + \frac{\alpha_2}{V_{gs}/V_{dd} + \alpha_3}\right)
    \end{equation}
    acts as a multi-variable correction whose functional form is suggestive of vertical field degradation and quantum mechanical subband splitting, though the fit to terminal current data cannot on its own distinguish this from other physical origins with similar functional dependence. Its saturating dependence on $V_{gs}$ is consistent with screening saturation of the vertical gate field, a behaviour that legacy linear or power-law formulations fail to track \cite{ando1982electronic, taur2021fundamentals}.
    
    \item \textbf{Native SPICE Syntax Compatibility:} A critical advantage of this discovered formulation is its compliance with compact modeling paradigms. Because the expression relies entirely on fundamental transcendental and algebraic operators (smooth square roots, natural logarithms, and basic arithmetic), it can be mapped directly into commercial circuit simulators via standard behavioral expressions (e.g., Verilog-A or \texttt{B-source} elements) without requiring look-up table approximations or introducing compilation performance penalties \cite{mcandrew2015bestpractices}.
\end{enumerate}

\section{SPICE Validation: DC and Transient Analysis}

\subsection{SPICE Implementation}
\label{sec:spice_impl}

Deriving a closed-form analytical expression is only valuable if it can be deployed in an actual circuit simulator without numerical failures. Two questions must be answered: (1) Does the recovered expression reproduce physically correct $I_D$--$V_{DS}$ and $I_D$--$V_{GS}$ characteristics across the full operating range? (2) Does the model remain numerically stable under dynamic switching conditions where voltages change rapidly and the SPICE Newton-Raphson solver must re-converge at each timestep?

We implement the recovered compact model in SPICE using a Behavioral Source (B-source) element \cite{brinson2016qucs}. Unlike standard device primitives, a B-source evaluates an arbitrary algebraic expression at runtime, making it the correct mechanism for prototyping new compact model equations \cite{chalkiadaki2013verilog} before committing them to a Verilog-A model card. The critical numerical requirements for B-source stability are:
\begin{itemize}
    \item All functions must be $\mathcal{C}^\infty$ continuous \cite{kundert1995spice} so that the SPICE Jacobian is always defined.
    \item No division by zero, no undefined logarithm arguments, and no discontinuous conditionals in the current path.
    \item The model must converge at DC operating point before any transient simulation can begin.
\end{itemize}

The $\text{smooth\_abs}$ operator introduced in Section~\ref{sec:regression} directly addresses all of these requirements. In SPICE syntax, it is implemented as:
\begin{lstlisting}
* mobility arguments in SI units: m^2 V^-1 s^-1
* alpha_0 = 1e-6 m^4 V^-2 s^-2
.func smooth_abs(x) {sqrt(x*x + 1e-6)}
\end{lstlisting}
This single function definition ensures that any absolute value operation in the model is differentiable at $x=0$, eliminating the non-differentiable cusp of the hard $|x|$ operator.

The full model is implemented as a Behavioral Source referencing the extracted closed-form expression for $\Delta\mu$ (Eq.~\ref{eq:final_symbolic_expression_formal}), combined with the analytical current integral equation from Section~\ref{sec:scaffolding}. Gate voltage, drain voltage, temperature, fin count, and geometry parameters are passed as node voltages or parameters.

\subsection{DC Family-of-Curves Analysis}

To validate basic device physics, we swept $V_{DS}$ from 0~V to 1.2~V at multiple fixed
$V_{GS}$ values, generating the $I_D$--$V_{DS}$ family of curves (output
characteristics). The MESD \texttt{N3A} training partition spans $0$--$0.7$~V, so the
$0$--$0.7$~V portion of the sweep is an in-domain check and the $0.7$--$1.2$~V portion is
an extrapolation test: it probes whether the expression remains smooth, monotonic and
convergent outside its fitted envelope, not whether it is quantitatively accurate there.
These curves are the canonical verification of MOSFET behavior:

\begin{itemize}
    \item \textbf{Below threshold ($V_{GS} < V_{th}$):} Curves lie near zero with very small current.
    \item \textbf{Linear region ($V_{DS} < V_{GS} - V_{th}$):} Current rises approximately linearly with $V_{DS}$, with slope increasing with $V_{GS} - V_{th}$.
    \item \textbf{Onset of saturation ($V_{DS} > V_{GS} - V_{th}$):} Current growth with $V_{DS}$ slows markedly beyond the knee, producing a compressive quasi-saturation shoulder rather than an exactly flat plateau.
    \item \textbf{Increasing $V_{GS}$:} Each successive curve carries more current, with well-separated, smooth trajectories.
\end{itemize}

\begin{figure}[H]
\centering
\includegraphics[width=0.7\textwidth]{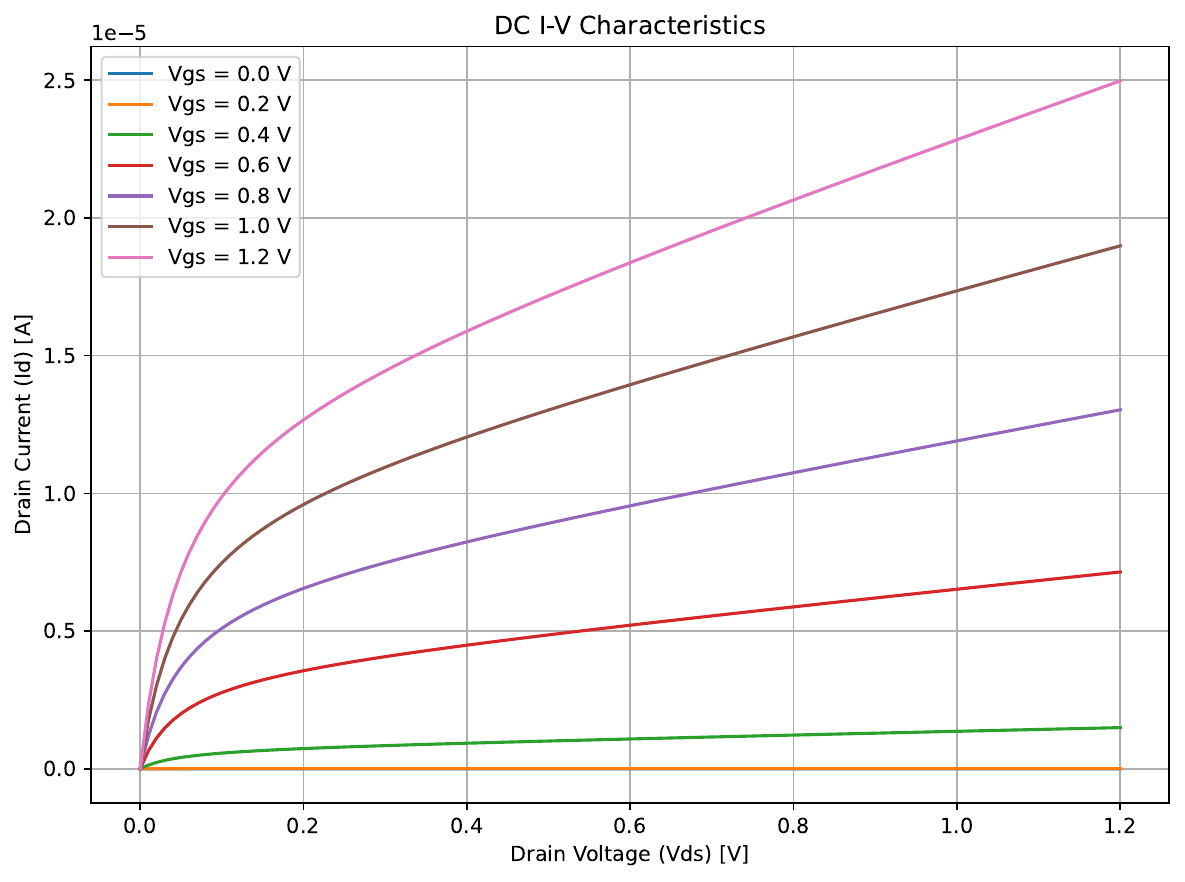}
\caption{Physical validity verification of the distilled symbolic model via SPICE DC family of curves ($I_D$--$V_{DS}$). The output characteristics are generated by implementing the recovered closed-form algebraic expression as a continuous behavioral source. Bias values up to $0.7$~V lie within the MESD \texttt{N3A} training envelope; the $0.7$--$1.2$~V region is extrapolation beyond the fitted domain and is included to test numerical robustness rather than quantitative accuracy. The model tracks the linear conduction region, exhibits a saturation knee that shifts with gate bias, and preserves exact zero-bias boundaries. The smooth, monotonic current ordering confirms that the physics-guided structure yields stable behavior across all simulated bias combinations.}
\label{fig:spice_dc}
\end{figure}

The SPICE simulation converged at all bias points without numerical failures, as shown
in Fig.~\ref{fig:spice_dc}. The recovered expression reproduces the subthreshold, linear
and quasi-saturated regions with smooth, well-separated curves that are monotonically
ordered by $V_{GS}$, and it remains continuous and convergent across the extrapolated
$0.7$--$1.2$~V region. Within the fitted envelope this confirms that the distilled
equation behaves as a physically valid device model rather than as an interpolation
artifact; beyond $0.7$~V the result establishes numerical robustness only, and
quantitative accuracy there is untested against reference data.

\subsection{Transient Square-Wave Switching Analysis}

The most demanding test of a compact model's SPICE compatibility is its behavior under
dynamic switching conditions. During a transient simulation, the gate voltage switches
rapidly between 0~V and 1.2~V with sharp edges (sub-nanosecond rise/fall times),
emulating a logic gate switching in a digital circuit. The upper level exceeds the
$0.7$~V MESD bias ceiling deliberately: the purpose of the test is solver convergence
under a step input that drives the expression outside its fitted domain, not accuracy of
the resulting current amplitude. This is substantially harder than DC analysis because:

\begin{itemize}[noitemsep]
    \item The SPICE solver must re-converge at each timestep as $V_{GS}$ sweeps rapidly through all operating regions.
    \item Any non-smooth or non-differentiable feature in the model causes the Newton-Raphson iteration to diverge.
    \item The model must handle simultaneous transitions in both $V_{GS}$ and $V_{DS}$ as the drain node responds dynamically to the current change.
\end{itemize}

\begin{figure}[H]
\centering
\includegraphics[width=0.80\textwidth]{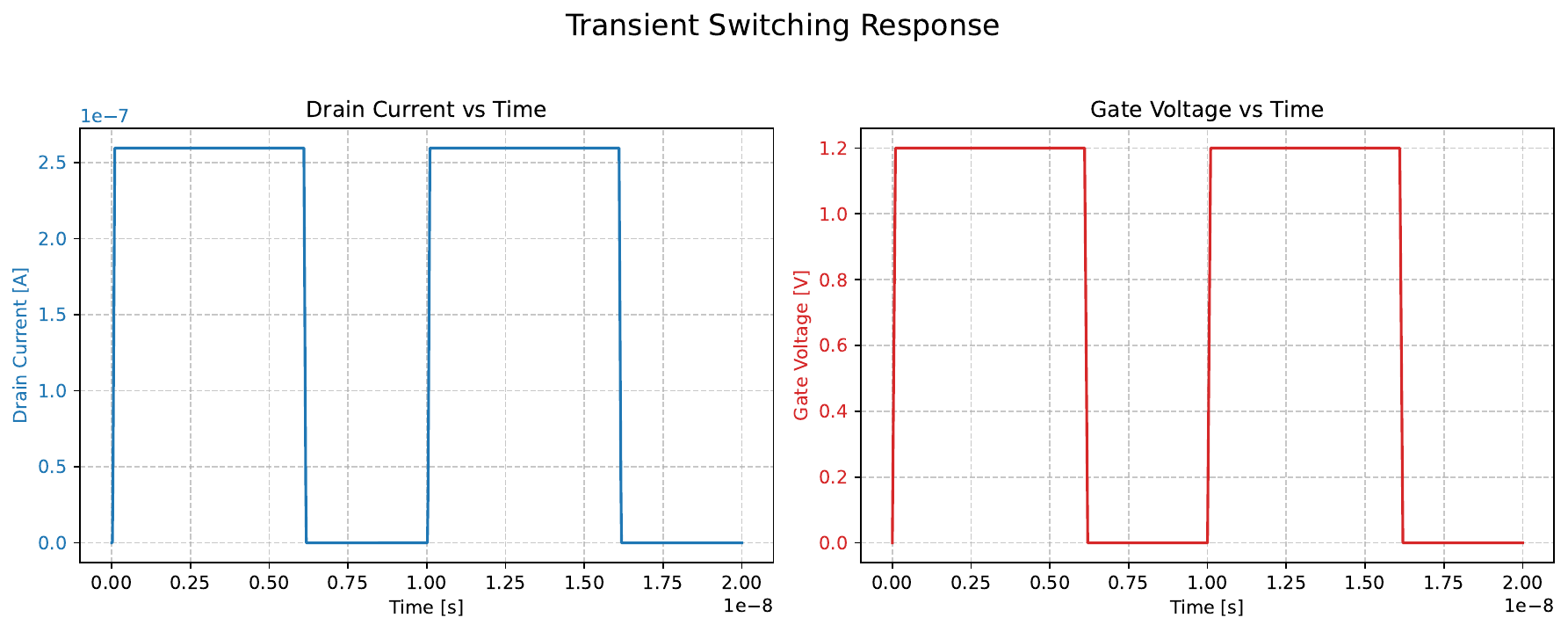}
\caption{Dynamic stability and numerical robustness verification of the distilled symbolic model via SPICE transient simulation. Left: The corresponding model-predicted channel current response ($I_D$). Right: The input waveform executing rapid square-wave gate voltage switching ($V_{GS}$) between 0~V and 1.2~V, the upper level lying beyond the $0.7$~V MESD training envelope. The clean current envelopes and absence of numerical oscillation under steep step-function inputs demonstrate that the distilled algebraic expression maintains numerical continuity and solver convergence, verifying its feasibility for stable transport-model prototyping inside circuit simulation engines. Current magnitudes above $0.7$~V are extrapolated and are not validated against reference data.}
\label{fig:spice_transient}
\end{figure}

The transient simulation completed successfully across all switching cycles without convergence failures, as shown in Fig.~\ref{fig:spice_transient}. The drain current responds cleanly to gate pulses, rising sharply when $V_{GS}$ exceeds threshold and falling back to the leakage floor when $V_{GS}$ returns to zero. The smooth transitions between on and off states, free of the numerical oscillations that indicate convergence issues, empirically confirm that the $\mathcal{C}^\infty$ smooth\_abs operator resolves all potential Newton-Raphson instabilities.

This result is particularly significant for industrial adoption. The ability to complete transient simulation without modification of the standard SPICE workflow means the recovered expression can be directly used in ring oscillator, logic gate, and standard cell simulation flows \cite{Weste2011cmos} without any special solver settings or convergence aids.

\section{Limitations}

While this framework successfully discovers compact analytical corrections for a 3nm FinFET PDK, several limitations remain. First, the current formulation focuses on steady-state DC transport mechanics and does not incorporate AC or transient characteristics such as parasitic capacitances ($C_{gs}$, $C_{gd}$) or gate-dielectric trapping kinetics. Extending the UDE framework to dynamic high-frequency behaviors would require optimizing over complex, time-dependent differential constraints. Second, the structural parameter extraction workflow relies on data from an established Process Design Kit (PDK). Evaluating physical hardware would introduce line-edge roughness, random dopant fluctuations, and measurement noise \cite{asenov1998variability}. Additional regularization constraints may be necessary to prevent the symbolic regression engine from overfitting noise profiles in physical measurement data \cite{sano2002rdf}. Finally, the symbolic expressions are locally optimal for the evaluated 3nm technology node. Translating this framework to Gate-All-Around (GAA) Nanosheets \cite{wang2024gaa} or Complementary FETs (CFETs) \cite{cfet2026modeling} would require modifying the base physical model equations to account for their unique electrostatic geometries.

\section{Conclusion and Future Work}

This work presented an automated pipeline for extracting closed-form analytical transport physics from 3\,nm FinFET simulation data and validating it directly inside a SPICE circuit simulation environment. Rather than hand-deriving a new equation for every physical effect, as is standard practice for compact models, we started from a structurally correct macroscopic transport model and used a Universal Differential Equation to let the data reveal what that baseline was missing. The resulting neural correction captured the device's geometry- and temperature-dependent transport behavior without requiring prior knowledge of the underlying microscopic mechanisms, and a two-stage optimization strategy combined with a multi-decade logarithmic loss allowed the framework to substantially reduce prediction error over the physical baseline while preserving physically meaningful subthreshold behavior. Symbolic regression then distilled this correction into the single closed-form expression given in Eq.~\eqref{eq:final_symbolic_expression_formal}, which reproduced the neural model's accuracy almost exactly and was validated directly in SPICE through both DC and transient switching analysis, confirming stable convergence and physically consistent behavior across all operating regions.

Several extensions would strengthen the practical reach of this framework. Incorporating dynamic charge-conservation effects ($C$--$V$ characteristics) alongside the transport equations studied here would allow the pipeline to support full transient circuit-level simulation rather than DC and switching behavior alone. Applying the same methodology to next-generation architectures such as Gate-All-Around nanosheets and Complementary FETs would test the generality of the UDE-to-symbolic-regression approach beyond the FinFET geometry studied here, and extending the framework to physical silicon measurements rather than PDK-simulated data would be a necessary step toward industrial deployment.

Taken together, these results suggest that combining physics-grounded baseline with data-driven residual learning and symbolic regression offers a practical, interpretable path toward automated compact model generation for future semiconductor technology nodes.

\section{Acknowledgments}
The authors acknowledge the use of AI tools during the preparation of this manuscript for assistance with rephrasing, paraphrasing text, and generating Fig. \ref{fig:methodology_overview}. All AI-generated content was thoroughly reviewed, verified, and edited by the authors, who take full responsibility for the accuracy and integrity of the work presented.

\bibliographystyle{unsrtnat}
\bibliography{references}

\end{document}